\documentclass[traditabstract]{aa} 
\usepackage{graphicx}
\usepackage{url}
\usepackage{natbib}
\usepackage{lscape}
\usepackage{longtable}
\usepackage{multirow}

\usepackage{txfonts}
\begin{document}
\newcommand{\Zsolar}{\mbox{$\,\rm Z_{\odot}$}}
\newcommand{\Msolar}{\mbox{$\,\rm M_{\odot}$}}
\newcommand{\Lsolar}{\mbox{$\,\rm L_{\odot}$}}
\newcommand{\xs}{$\chi^{2}$}
\newcommand{\dxs}{$\Delta\chi^{2}$}
\newcommand{\xsn}{$\chi^{2}_{\nu}$}
\newcommand{\ls}{{\tiny \( \stackrel{<}{\sim}\)}}
\newcommand{\gs}{{\tiny \( \stackrel{>}{\sim}\)}}
\newcommand{\asec}{$^{\prime\prime}$}
\newcommand{\amin}{$^{\prime}$}
\newcommand{\mstar}{\mbox{$M_{*}$}}
\newcommand{\hi}{H{\sc i}}
\newcommand{\hii}{H{\sc ii}\ }
\newcommand{\kms}{$\rm km~s^{-1}$}

   \title{The GALEX view of the {\it Herschel} Reference Survey}
   \subtitle{Ultraviolet structural properties of nearby galaxies} 

\author{
L. Cortese\inst{1}
\and
S. Boissier\inst{2}
\and
A. Boselli\inst{2}
\and
G. J. Bendo\inst{3}
\and 
V. Buat\inst{2}
\and
J. I. Davies\inst{4}
\and
S. Eales\inst{4}
\and
S. Heinis\inst{2}
\and
K. G. Isaak\inst{5}
\and
S. C. Madden\inst{6}
}

\institute{
European Southern Observatory, Karl Schwarzschild Str. 2, 85748 Garching bei M\"unchen, Germany
\and
Laboratoire d'Astrophysique de Marseille - LAM, Universit\'e d'Aix-Marseille \& CNRS, UMR7326, 38 rue F. Joliot-Curie, 13388 Marseille Cedex 13, France
\and
UK ALMA Regional Centre Node, Jodrell Bank Centre for Astrophysics, School of Physics and Astronomy, University of Manchester, Oxford Road,
Manchester M13 9PL, United Kingdom
\and
School of Physics and Astronomy, Cardiff University, The Parade, Cardiff, CF24 3AA, UK
\and
European Space \& Technology Centre (ESTEC), PO Box 299, 2200 AG Noordwijk, the Netherlands
\and
Institut d'Astrophysique Spatiale (IAS), Batiment 121, Universite Paris-Sud 11 and CNRS, F-91405 Orsay, France 
}

   \date{Received 30 March 2012 - Accepted 30 May 2012}

 
  \abstract{We present GALEX far-ultraviolet (FUV) and near-ultraviolet (NUV) as well as SDSS $g$, $r$, $i$ photometry 
  and structural parameters for the {\it Herschel} Reference Survey, a magnitude-, volume-limited sample of nearby 
  galaxies in different environments. We use this unique dataset to investigate the ultraviolet (UV) structural scaling relations 
  of nearby galaxies and to determine how the properties of the UV 
  disk vary with atomic hydrogen content and environment. We find a clear change of slope in the stellar mass vs. 
  effective surface brightness relation when moving from the optical to the UV, with more massive galaxies 
  having brighter optical but fainter UV surface brightnesses than smaller systems. A similar change of slope is also 
  seen in the radius vs. surface brightness relation. By comparing our observations with the predictions of a simple 
  multi-zone chemical model of galaxy evolution, we show that these findings are a natural consequence of a much more 
  efficient inside-out growth of the stellar disk in massive galaxies. 
  
  We confirm that isophotal radii are always a better proxy for the size of the stellar/star-forming disk than effective 
  quantities and we show that the extent of the UV disk (normalized to the optical size) is strongly correlated to the integrated 
  \hi\ gas fraction. This relation still holds even when cluster spirals are considered, with \hi-deficient 
  systems having less extended star-forming disks than \hi-normal galaxies. Interestingly, the star formation in the inner 
  part of \hi-deficient galaxies is significantly less affected by the removal of the atomic hydrogen, 
  as expected in a simple ram-pressure stripping scenario. These results suggest that it is the amount of \hi\ that regulates 
  the growth of the star-forming disk in the outskirts of galaxies. 
  
  }


	\authorrunning{Cortese et al.}	
	
   \maketitle
%

\section{Introduction}
Nearly a decade has passed since the launch of the {\it Galaxy Evolution Explorer} (GALEX; \citealp{martin05}) 
in 2003. By carrying out the first all-sky survey at ultraviolet (UV) wavelengths, GALEX has 
opened a new parameter space for extragalactic studies, providing a direct estimate of the current 
star formation rate in relatively dust-free environments. 
Particularly powerful for our understanding of star formation 
in galaxies has been the combination of GALEX observations with multiwavelength datasets 
tracing dust-obscured star formation (e.g., \citealp{sings}) and the various components of the interstellar medium (ISM; e.g., \citealp{bigiel08}).
Consequently, UV observations have now become a necessary ingredient of any multiwavelength survey focused on the study 
of the physical mechanisms regulating the star formation cycle in galaxies. 

In this paper, we present GALEX observations of the galaxies in the {\it Herschel} Reference Survey (HRS, \citealp{HRS}), 
a {\it Herschel} guaranteed time project focused on the study of the interplay between dust, gas and star formation 
in the local Universe. This dataset complements the Local Volume GALEX survey \citep{lee11}, since 
it includes more massive systems and it covers a wider range of environments. Thus, once combined, these two surveys 
provide us with a complete view of UV properties of galaxies within $\sim$25 Mpc from us. 

Given the large apparent size of the galaxies in the HRS, GALEX observations are particularly 
suitable to investigate the UV morphology of nearby galaxies and their connection to internal 
galaxy properties and environment. 
Indeed, still very little is known about the UV structural properties of the local galaxy population.
 
The first systematic investigation of the GALEX UV morphology in nearby galaxies has been presented in the seminal work carried out 
by \cite{atlas2006}. Although this analysis has been followed by several studies focused 
on colour gradients (e.g., \citealp{mateos07}), UV extended disks (e.g., \citealp{thilker07,lemonias11}), 
ellipticals (e.g., \citealp{jeong09}) and dwarf galaxies (e.g., \citealp{zhang12}), we are still missing 
an accurate quantification of the UV structural scaling relations of nearby galaxies.
Firstly, it is still unknown whether structural scaling relations such as the stellar mass vs. size, stellar mass vs. 
surface brightness and size vs. surface brightness, which represent important constraints for theoretical models, hold also at 
UV wavelengths. Since UV is the ideal tracer of current star formation in the unobscured outer parts of galaxies, 
the UV scaling relations can provide us with strong constraints on the growth of the stellar disk and on the 
origin of the extended UV-disk phenomenon. Secondly, while several studies have highlighted how well UV traces the \hi\ content 
in the outer part of galaxies \citep{bigiel10,bigiel10b,lemonias11}, it still has to be proven that such a tight relation 
holds for the whole galaxy population. Thirdly, it is still unclear how the cluster environment affects the UV morphology. 
Although previous studies have shown that in \hi-deficient/cluster galaxies the extent of the H$\alpha$ disk 
is significantly reduced compared to those in \hi-normal/field systems \citep{koop1,koop2,koop06,review}, 
it is still unclear if the UV disks are truncated as well and whether or not the \hi\ stripping affects the star formation in the 
central regions of stripped galaxies \citep{review}. 

For all these reasons, in this paper we investigate the UV structural properties of galaxies in the HRS in order to determine 
how they vary with galaxy properties, gas content and environment. As a by product of this analysis we present, in addition 
to the GALEX data, optical photometry and structural parameters in the SDSS $g$, $r$ and $i$ bands.
 
This work is part of our current effort to make publicly available to the astronomical community all the 
multiwavelength datasets collected for the HRS (e.g., \citealp{bendo12,boselli12b,ciesla12,hughes12}), 
thus enhancing the legacy value of this survey in the years to come.


\section{The data}
\subsection{Sample Selection}
The HRS is a volume-limited sample 
(i.e., 15$\leq Dist. \leq$25 Mpc) including late-type galaxies (Sa and later) with  2MASS \citep{2massall} 
K-band magnitude K$_{Stot} \le$ 12 mag and early-type galaxies (S0a and earlier) with K$_{Stot} \le$ 8.7 mag. 
Additional selection criteria are high galactic latitude ($b>$ +55$^{\circ}$) and low Galactic extinction 
($A_{B}$ $<$ 0.2 mag, \citealp{schlegel98}), to minimize Galactic cirrus contamination. 
The original selection included 323 galaxies (261 late- and 62 early-types), although later 
one galaxy (HRS228) was excluded due to a wrong redshift reported in NED (see also \citealp{cortese12}). 
 
\subsection{GALEX Observations}
In order to obtain homogeneous near-ultraviolet 
(NUV; $\lambda$=2316 \AA: $\Delta \lambda$=1069 \AA) and far-ultraviolet 
(FUV; $\lambda$=1539 \AA: $\Delta \lambda$=442 \AA) data with an exposure time of at least 
$\sim$1.5 ksec (corresponding to a surface brightness limit of $\sim$28.5 mag arcsec$^{-2}$) 
for all the galaxies in the HRS observable by GALEX, we were 
awarded 112.5 ksec as part of the legacy GI Cycle 6 proposal {\it Completing 
the GALEX coverage of the Herschel Reference Survey} (P.I. L. Cortese). 
Unfortunately, before the start of Cycle 6 observations, the FUV detector 
experienced an over-current condition and shut down, and 
GALEX officially moved to NUV-only operations. 
In addition, subsequent problems to the NUV detector and budget cuts on the 
mission did not allow GALEX to complete the GI and MIS surveys as planned.

For all these reasons, the GALEX coverage of the HRS remains heterogeneous 
and the observations here presented come from a combination of our GI proposal 
with data from the GALEX Ultraviolet Virgo Cluster Survey (GUViCS, \citealp{guvics}) and archival observations 
publicly available as part of the GALEX GR6 data release. 
In detail, NUV observations are available for all HRS galaxies 
observable by GALEX (310 galaxies\footnote{13 galaxies cannot be observed because too close 
to bright stars exceeding the counts rate allowed by the NUV detector.}): 285 galaxies 
have been observed with exposure time longer than 1ksec (82 from our proposal and 
the rest as part of the Nearby Galaxy Survey, Medium Imaging Survey and other Guest Investigator 
programs), while the remaining 26 galaxies have a typical integration time of $\sim$200 sec 
and come mainly from the All Sky Imaging Survey (AIS). FUV observations are available for 302 galaxies: 
167 with  exposure times longer than 1ksec and the rest coming from the AIS. However, for 29 galaxies 
the AIS tiles were either too shallow to detect the target or the galaxy was on the edge of the field 
making impossible to perform reliable photometry. Thus, in this paper we present FUV magnitudes and 
structural parameters for just 273 galaxies ($\sim$85\% of the HRS). 
All frames have been reduced using the current version of the GALEX pipeline (ops-v7\footnote{\url{http://galex.stsci.edu/doc/GI_Doc_Ops7.pdf}}).  
Details about the GALEX satellite and data reduction can be found in \cite{martin05} and \cite{morrissey07}.

Table~\ref{tab:hrsgalaxies} lists some general properties of the HRS galaxies, the 
GALEX tiles and corresponding exposure times used in this work. 

\subsection{SDSS optical data}
We combine the GALEX data with $g$, $r$ and $i$ band images for the 313 HRS galaxies 
included in the {\it Sloan Digital Sky Survey} DR7 (SDSS-DR7, \citealp{sdssDR7}) footprint. 
For those cases where our target was present in more than one SDSS frame, we used the \textsc{imcombine} task 
in IRAF\footnote{IRAF is distributed by the National Optical Astronomy Observatory, which is operated by 
the Association of Universities for Research in Astronomy (AURA) under cooperative agreement with the National Science Foundation.} 
to create a mosaic and recover all the emission from the galaxy at least up to the optical radius.   
\begin{figure*}
  \centering
  \includegraphics[width=16cm]{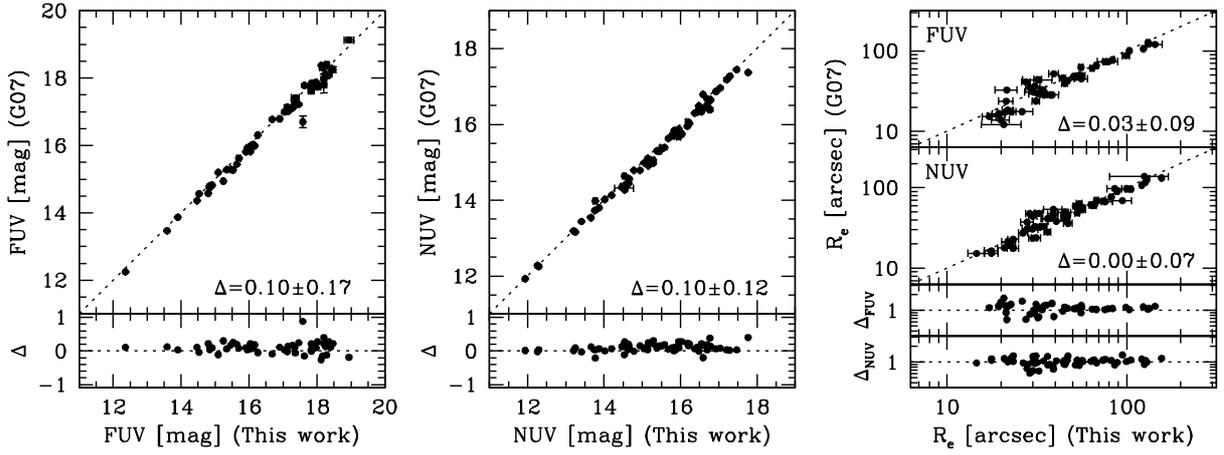}
     \caption{Comparison of the FUV (left) and NUV (center) magnitudes and effective radii (right) presented here with the ones obtained by 
     Gil de Paz et al. 2007 (G07). The average differences (i.e., this work - G07) and standard deviations are shown in each panel. The dotted lines indicate the 1-to-1 relation.}
	 \label{check}
  \end{figure*}

\section{Photometry}
Both the GALEX and SDSS pipelines are not optimized for extended sources, thus 
we performed our own photometry starting from the reduced and calibrated frames. 
The SDSS and FUV images were registered to the NUV frame using the \textsc{wcsmap} and \textsc{geotran} tasks in IRAF  
and convolved to the NUV resolution (5.3\arcsec, \citealp{morrissey07}). 
In the few cases for which GALEX images were not available, we just re-binned the SDSS frames to the same 
pixel size of GALEX data (1.5 arcsec) and convolved them to the NUV resolution.

Surface brightness photometry was performed using a modified version of the GALPHOT \citep{haynesgalphot} IRAF package, 
adapted in order to work on GALEX data.
Sky background was determined in rectangular regions around the target, chosen 
independently in each band to avoid background/foreground sources, artifacts and the emission from the target. 
The mean sky value was then subtracted from each frame. 
Background/foreground sources and artifacts were then masked in each image independently and a final mask 
was created by merging all the pixels masked in the different bands.
Isophotal ellipses were fitted to each sky-subtracted image by using the IRAF task \textsc{ellipse}. 
When available, the SDSS $i$-band frame was used to define the galaxy center, 
ellipticity and position angle. Otherwise, we used either the ellipticity and position angle listed in the RC3 catalogue \citep{rc3} or 
determined it directly from the images available. Center, position angle and ellipticity were then kept fixed. The fitting always started 
with the most central ellipse having a major axis radius of 6 arcsec, increasing linearly by 3 arcsec for each step.   
Uncertainties on surface brightnesses and integrated magnitudes within each ellipse were determined 
following \cite{atlas2006} and \cite{bosiso03} (see also \citealp{mateos09a}).
Asymptotic magnitudes in each band have been estimated through a linear weighted fit of the growth curve, 
following the technique described by \cite{atlas2006} (see also \citealp{cairos01}). In the rest of the paper, we will consider the 
asymptotic magnitudes as our best estimate for the total galaxy flux in each band. All magnitudes are given in the AB system.
 
It is important to note that the estimated uncertainties in the magnitudes are a combination of 
the error on the sky background determination, the Poisson error on the incident flux and the calibration error (0.05, 0.03, 
0.03, 0.02, 0.03 mag in FUV, NUV, $g$, $r$, $i$, respectively; \citealp{morrissey07,sdssDR7}). 
Thus, they do not take into account possible additional flat-field variations across the frame (affecting mainly extended sources) 
and the fact that shallow ($\lesssim$ 200 sec) GALEX images are not background- but source-limited. Indeed, by comparing independent 
GALEX observations of the same target, we find a standard deviation in the recovered asymptotic magnitude 
of $\sim$0.1-0.15 mag, i.e., larger than the typical errors obtained following the standard procedure 
described above (i.e. $\sim$0.06 in FUV and 0.04 NUV, see also \S~3.1). 

From the surface brightness profiles, we also determined integrated magnitudes within the optical diameters 
given in Table~\ref{tab:hrsgalaxies}, effective radii ($R_{e}$, i.e. the radius containing 50\% of the total light), 
effective surface brightnesses ($<\mu_{e}>$, i.e., the average surface brightness inside $R_{e}$) and isophotal radii. 
The isophotal radii have been computed at surface brightness levels of 23.5, 24, 24.5, 28 and 28 mag arcsec$^{-2}$ in 
$i$, $r$, $g$, NUV and FUV, respectively. These values roughly correspond to the average surface brightness at the 
optical diameter observed for the whole sample. Asymptotic magnitudes are on average $\sim$0.1$\pm$0.1 mag brighter 
than those estimated up to the optical diameter.
The typical uncertainty in the effective and isophotal radii varies between $\sim$20\% in FUV to $\lesssim$10\% in the other bands.
Similarly, the error on the effective surface brightness increases from $\lesssim$0.1 mag for the SDSS and NUV bands to $\sim$0.20 mag  
in FUV. The larger uncertainty in the FUV structural parameters is simply due to the fact that nearly half of the 
FUV photometry has been performed on shallow AIS frames. This must be taken into account when a comparison between the scaling relations 
in the two GALEX bands is performed. 

Finally, we note that we did not apply any corrections for inclination to our photometry and structural parameters, since 
they are usually highly uncertain (e.g., \citealp{giovanelli94,giovanelli95}). Although our approach may artificially increase 
the scatter in some of the scaling relations investigated in the rest of the paper (in particular the ones involving radii), the main conclusions 
of our work are not affected. 

The results of our photometry (not corrected for Galactic extinction) are presented in Table \ref{tab:hrsmags} and \ref{tab:struct}.
Notes on individual problematic or peculiar objects are given in Appendix A.

The results of our photometry as well as the GALEX data are publicly available on the Herschel Database in Marseille (HeDaM, \url{http://hedam.oamp.fr/}).
\begin{figure*}
  \centering
  \includegraphics[width=15.5cm]{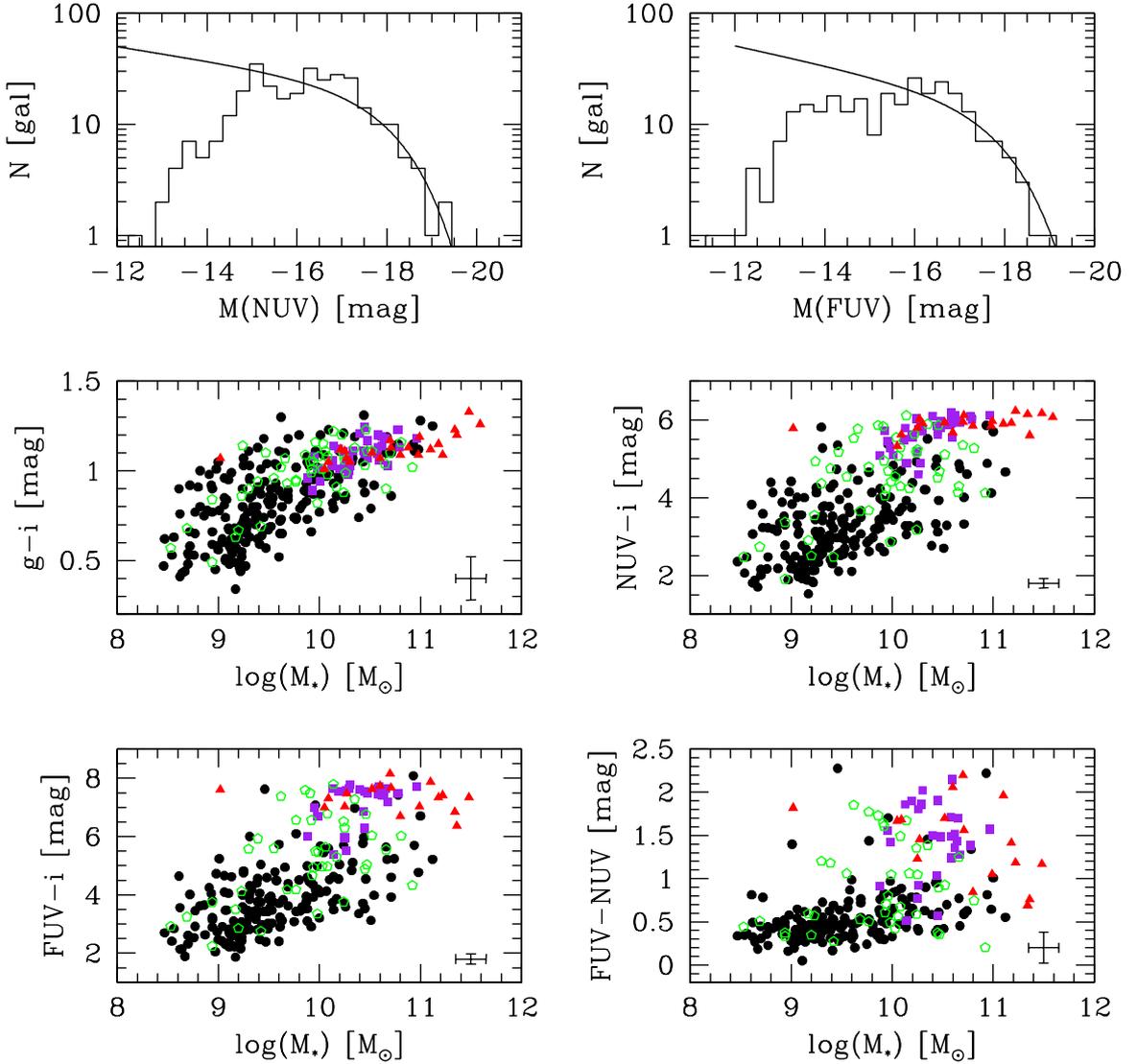}
     \caption{The basic UV and optical properties of the HRS. Top row: The NUV (left) and FUV (right) magnitude distribution. 
     The solid line shows the local GALEX UV luminosity function presented by \cite{wider}. Middle row: $g-i$ (left) and 
     $NUV-i$ (right) colour vs. stellar mass relations. Bottom row: $FUV-i$ (left) and 
     $FUV-NUV$ (right) colour vs. stellar mass relations. Colours are corrected for Galactic extinction only. Galaxies are colour-coded accordingly 
     to their morphological type: red triangles are E-dE, purple squares S0-S0a, green pentagons Sa-Sab and black circles Sb and later types.}
	 \label{basic}
  \end{figure*}

\subsection{Comparison with the literature}
In order to check the reliability of the GALEX measurements presented here, we compare our UV asymptotic magnitudes 
with the values obtained by \cite{atlas2006}. This is the only UV catalogue currently 
available with significant overlap with the HRS: i.e., 62 and 52 galaxies in NUV and FUV, respectively. 
The results of this comparison are shown in Fig.~\ref{check}. Overall there is good agreement between 
the two studies, with a typical scatter of $\sim$0.12 mag in NUV, $\sim$0.17 mag in FUV magnitudes and $\sim$0.08 dex in 
effective radius. However, we find a systematic offset between the two compilations, with our fluxes being 
$\sim$0.1 mag fainter in both NUV and FUV than \cite{atlas2006}. This is most likely due to 
the change in flux calibration since the GALEX GR1 release (used by \citealp{atlas2006}), as 
also suggested by the fact that no systematic offset is seen when the effective radii are considered.

It is well known that the automatic SDSS photometry pipeline is not reliable for extended sources, such 
as the HRS sample. This is mainly due to problems in background subtraction and blending of large galaxies into multiple 
sources \citep{bernardi07,west10}. Indeed, a comparison between our asymptotic magnitudes and the cmodel SDSS magnitudes given in NED shows an average 
difference of $\sim-$0.8$\pm$1.0 mag, with our estimates being brighter. Luckily, some galaxies in our sample have already been 
remeasured as part of previous studies and we compared our asymptotic magnitudes with the published values. 
We find an average difference between our values and those published of $\sim$0.00$\pm$0.05, $\sim$0.08$\pm$0.12 and $-$0.01$\pm$0.08 mag 
for the samples of \cite{mateos09a}, \cite{mcdonald11} and \cite{chen10}\footnote{In this case, we considered the values obtained from the growth curve 
analysis since much more similar to the 
technique adopted here.}, respectively. These small differences are likely due to the different techniques used to determine total magnitudes. 
We can thus assume a typical uncertainty of $\sim$0.1 mag in the SDSS magnitudes presented in Table \ref{tab:hrsmags}.

\section{Basic properties}
Fig. ~\ref{basic} provides a general overview of the UV and optical properties of the HRS. 
The top row shows the NUV and FUV luminosity distributions (corrected for Galactic extinction following \citealp{wyder07}) 
for all the galaxies for which GALEX observations are available. For comparison, 
the local UV GALEX luminosity function of \cite{wider} is presented. This has been arbitrarily normalized to match the bright-end of 
the luminosity distribution of the HRS. Our sample turns out to be a good representation of the UV luminosity distribution in the local 
Universe up to M(NUV)$\sim-$15 and M(FUV)$\sim-$16 mag, whereas at lower luminosities we under-sample the population of faint UV sources. 
In addition to the well known Malmquist bias, this result is due to the HRS being a K-band selected sample, thus missing low-mass 
star-forming galaxies. 

The middle and bottom rows of Fig.~\ref{basic} highlight the different behavior of our sample in different colour vs. stellar mass relations. 
As already shown by several works (e.g., \citealp{bosell05,atlas2006,wyder07,cortese09,cortese12b}), the UV-to-optical colours are much more powerful 
than the optical-only colours in separating the blue cloud from the red sequence. Only with UV magnitudes, a colour cut is 
almost as effective as a morphological classification in separating early- and late-type galaxies, even before any 
correction for internal dust attenuation. 
It is also interesting to note how the various morphological types behave when moving from an UV-to-optical to a FUV-NUV colour vs. stellar mass 
diagram. The blue cloud becomes almost a blue sequence, while for early-type galaxies the scatter significantly increases, and galaxies 
are dispersed across a range of more than 2 mag in colour. Finally, we remind the reader that the absence of red sequence galaxies 
for stellar masses lower than 10$^{10}$ M$_{\odot}$ is a consequence of the different K-band magnitude selection used for early- and late-type galaxies. 
This can be easily seen by comparing Fig.~\ref{basic} with Fig. 1 of \cite{hughes09}, who used the same K-band magnitude cut 
for all galaxies, regardless of their morphology.

\section{Structural scaling relations}

\begin{figure*}
  \centering
  \includegraphics[width=18cm]{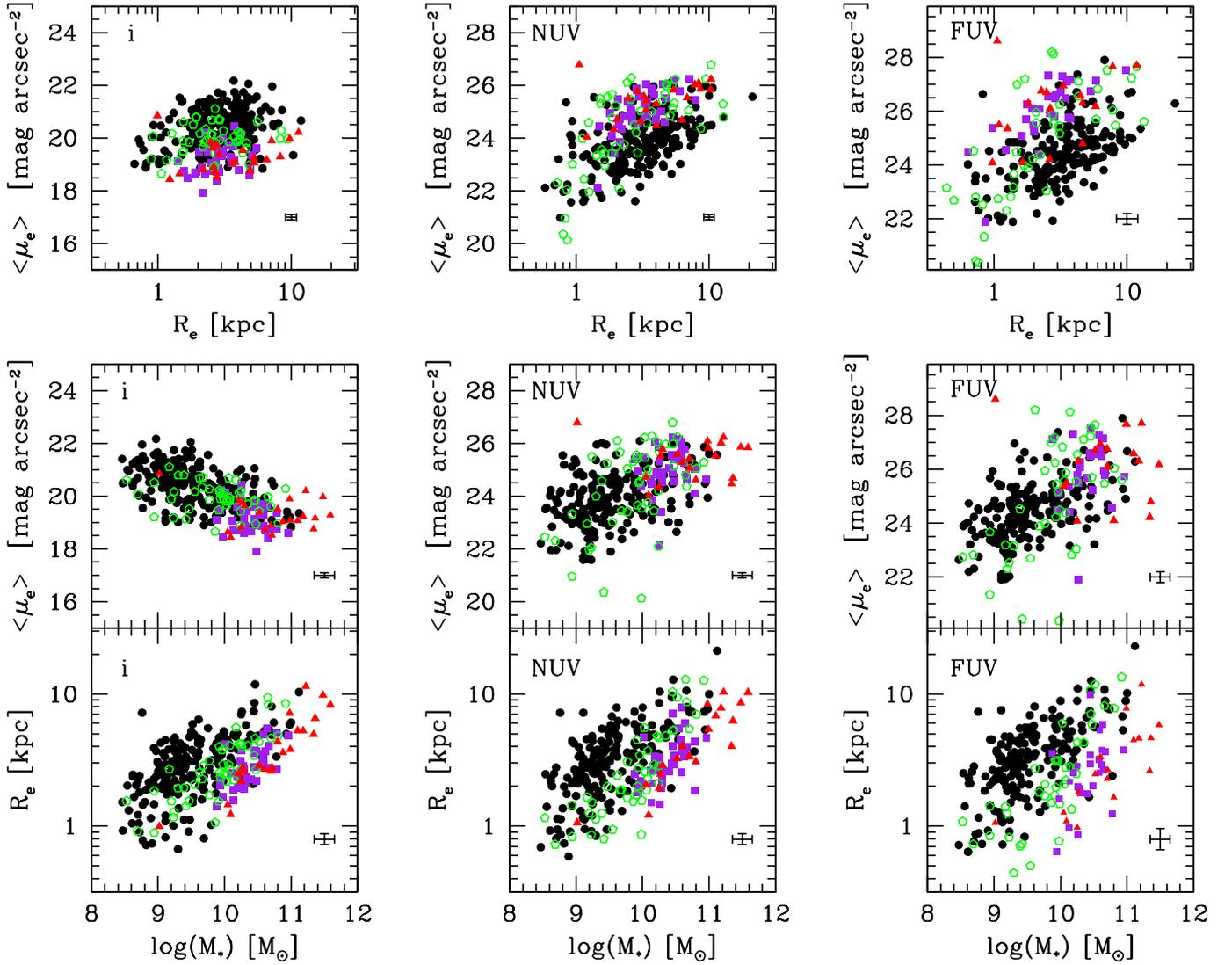}
     \caption{The effective surface brightness vs. effective radius (top row), effective surface brightness vs. stellar mass (middle row) 
     and stellar mass vs. effective radius (bottom row) relations in $i$ (left), NUV (center) and FUV (right), respectively. Symbols are as 
     in Fig.~\ref{basic}.}
	 \label{kormendy}
  \end{figure*}
\begin{figure*}
  \centering
  \includegraphics[width=18cm]{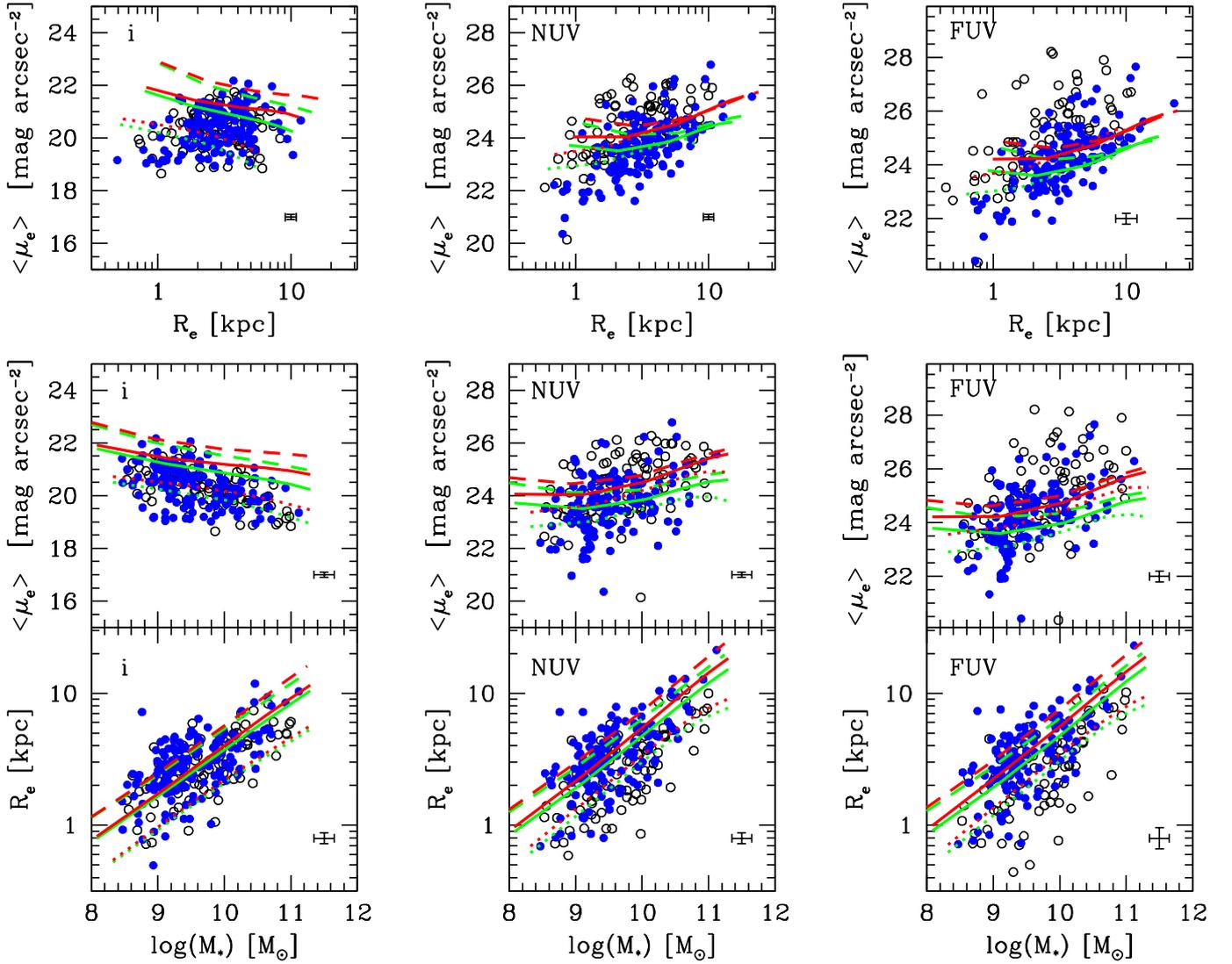}
     \caption{Same as Fig.~\ref{kormendy}, but for \hi-normal (filled circles) and \hi-deficient (empty circles) spirals (Sa and later types). 
     The predictions of the dust-free models by \cite{boissier00} for spin parameter 0.02 (dotted), 0.04 (solid) and 0.06 (dashed) 
     are superposed in green. The red lines indicate the same predictions once the effect of dust attenuation is included. }
	 \label{kormendylate}
  \end{figure*}
The structural parameters presented in Table~\ref{tab:struct} allow us to investigate, for the first time, 
how the scaling relations between stellar mass ($M_{*}$), effective radius ($R_{e}$) and effective 
surface brightness ($<\mu_{e}>$, corrected for Galactic extinction) behave when UV bands are considered here. 
The main results of this analysis are shown in Figs.~\ref{kormendy} and \ref{kormendylate}, 
where the effective radii vs. stellar mass, effective surface brightness vs. stellar mass and effective 
radii vs. effective surface brightness relations in $i$, NUV and FUV are shown in the left, central and right 
column, respectively. Galaxies are separated accordingly to their morphological type in Fig.~\ref{kormendy} and 
to their \hi\ content in Fig.~\ref{kormendylate}.
We computed the \hi-deficiency\footnote{The 
H{\sc i}-deficiency ($Def_{HI}$) is defined as the difference, in 
logarithmic units, between the expected H{\sc i} mass for an isolated galaxy 
with the same morphological type and optical diameter of the target and the observed value \citep{haynes}.} 
($Def_{HI}$) parameter for the HRS galaxies following \cite{cortese11}.
Atomic hydrogen masses have been estimated from \hi~21 cm line emission data (mainly single-dish), available from 
the literature (e.g., \citealp{spring05hi,giovanelli07,kent08,goldmine} and the {\it NASA/IPAC Extragalactic Database},
NED). We adopted  a threshold of $Def_{HI}=$0.5 to separate \hi-normal (filled blue circles) from \hi-deficient 
galaxies (empty circles). We assumed
\begin{equation}
\frac{M(HI)}{M_{\odot}}= 2.356 \times 10^{5} \frac{S_{HI}}{\rm Jy~km s^{-1}} \Big(\frac{Dist.}{\rm Mpc}\Big)^{2} 
\end{equation}
where $S_{HI}$ is the integrated \hi~line flux-density. 
In Table~\ref{scaletab}, we present the Pearson correlation and bisector\footnote{We decided to use the bisector fit in order 
to provide a direct comparison with theoretical models, but our conclusions do not qualitatively change 
if other linear regression techniques are adopted. We remind the reader that, by construction, the bisector best-fit must 
not be used to predict one quantity from the other \citep{linear}.} best linear fit coefficients \citep{linear} 
for all the three scaling relations considered here. 

Starting from the bottom panels of Fig.\ref{kormendy}, the $M_{*}$ vs. $R_{e}$ does not show any strong wavelength dependence 
and the effective radii always increase monotonically with stellar mass. Interestingly, at both optical and UV wavelengths, 
different morphological types follow slightly different relations, with early-type systems having smaller sizes 
(i.e., being more compact) than late type systems, at fixed stellar mass (e.g., \citealp{scodeggio02,shen03}). 
No offset is found between \hi-normal and \hi-deficient galaxies (see Fig.~\ref{kormendylate}) in the optical, whereas in UV 
\hi-deficient galaxies seem to have smaller radii than \hi-normal systems. However, this difference is marginally significant. 
As already mentioned in \S~3, at this stage, it is impossible to determine whether the larger scatter observed for early-type galaxies 
in FUV is real or it is just a consequence of the larger errors in the estimate of the FUV structural parameters. 

The most remarkable difference between UV and optical structural scaling relations is clearly seen in the 
$M_{*}$ vs. $<\mu_{e}>$ relation. While in $i$-band the effective surface brightness becomes brighter at increasing stellar 
mass, the opposite trend is seen in UV. A similar result was found by \cite{phenomen}, who investigated the relation 
between UV, B- and H-band surface brightness (normalized to the optical radius) and H-band luminosity for a large 
sample of cluster spirals. This is the first time that such trend is confirmed with GALEX UV data, and for a sample 
covering the whole range of morphologies. Our result implies that, at least in the local Universe, a UV-selected sample 
would preferentially be biased towards low-mass, actively star-forming galaxies, missing the more massive systems.    

In order to determine the origin of such remarkable inversion in the 
$M_{*}$ vs. $<\mu_{e}>$ relation, it is important to examine separately different morphological types. Indeed, while in 
spirals it is fair to assume that most of the UV photons come from young stellar populations, this is not 
the case for early-type galaxies where NUV and FUV fluxes are likely contaminated by more evolved stellar 
populations (e.g., \citealp{connell,bosell05,donas06,han07}). If we focus our attention on E and S0 only (red triangles and purple 
squares in Fig.~\ref{kormendy}), we no longer see 
an inversion in the $M_{*}$ vs. $<\mu_{e}>$ relation. At all wavelengths, the two quantities 
are only weakly correlated (see also Table~\ref{scaletab}) and the only significant difference between the optical and UV relations is that early-type 
systems are the objects with the brightest optical and faintest UV surface brightness in our sample. This is just a natural consequence of the 
fact that the luminosity of early-type galaxies decreases 
by $\sim$3 orders of magnitude when moving from the optical to the UV regime, but the effective radius stays almost the same, 
explaining why the effective surface brightness decreases so remarkably. 

Much more intriguing is the relation between $M_{*}$ and $<\mu_{e}>$ for late-type galaxies. Here, 
we always find an inverse correlation between $M_{*}$ and $<\mu_{e}>$ in optical, and a direct correlation in the UV, regardless of the 
criteria used to divide our sample (e.g., by morphology or gas content\footnote{Even in this case, 
at fixed stellar mass, \hi-deficient galaxies seem to show marginally fainter $<\mu_{e}>$ compared to \hi-normal galaxies.}, see Table~\ref{scaletab} 
and Figs.~\ref{kormendy} and \ref{kormendylate}). 
The fact that more massive disks have brighter effective surface brightness is a well known property of late-type spirals \citep{phenomen,scodeggio02} 
and it is usually interpreted as a consequence of the fact that massive disks have built up their stellar mass at earlier epochs  
than smaller systems following an inside-out growth of the stellar disk (e.g., \citealp{dalcanton97,boissier00,dutton09}). 
Thus, massive galaxies have already consumed a significant fraction of their gas reservoir in the center and their star formation 
surface density is lower than in dwarf systems. 
In other words, this is just a consequence of the anti-correlation between specific star formation rate and 
stellar mass (e.g., \citealp{salim07}).
Of course, it is important to remember that 
the surface brightnesses values shown in  Figs.~\ref{kormendy} and \ref{kormendylate} are not corrected for internal dust attenuation, 
and thus part of this trend might just be a consequence of the fact that more massive star-forming systems are more affected 
by dust than dwarf galaxies. However, in our mass range, the dependence of extinction on stellar mass is quite weak, 
and it could introduce a spurious systematic trend of $\sim$1-1.5 mag (e.g., \citealp{cortese06,jorge06}): i.e., 
enough to flatten the relation, but insufficient to explain the reversal of the $M_{*}$ vs. $<\mu_{e}>$ in UV.   

In order to test this interpretation, we compared our observations with the predictions of the multi-zone chemical and 
spectrophotometric model of \cite{boissier00}, updated with an empirically determined star formation law \citep{boissier03} 
relating the star formation rate to the total gas surface density. The only two free parameters in this model 
are the spin parameter $\lambda$ (e.g., \citealp{mo98}), and the rotational velocity, $V_{C}$. 
The star formation history of a galaxy depends on the infall timescale, which is a function of $V_{C}$. Thus, 
$V_{C}$ controls the stellar mass accumulated during the history of the galaxy, and $\lambda$ its radial distribution.
We assumed an age of 13.5 Gyr, varied the spin parameter $\lambda$ from 0.02 to 0.09 (with 0.01 step) and considered 
five different values for $V_{C}=$40, 80, 150, 220, 290, 360 km s$^{-1}$. 
We adopted both the dust-free model and the reddened 
one obtained as described in \cite{boissier99}. Finally, the model stellar masses have been converted from a 
\cite{kroupa93} IMF, used in the model, to a \cite{chabrier} IMF by adding 
0.06 dex \citep{bell03,gallazzi08}. 
This model is able to reproduce the integrated properties \citep{boissier00} as well as the surface brightness profiles \citep{munoz11} 
of nearby late-type galaxies, thus it is an ideal tool for a comparison with our scaling relations.   
\begin{figure*}
  \centering
  \includegraphics[width=17cm]{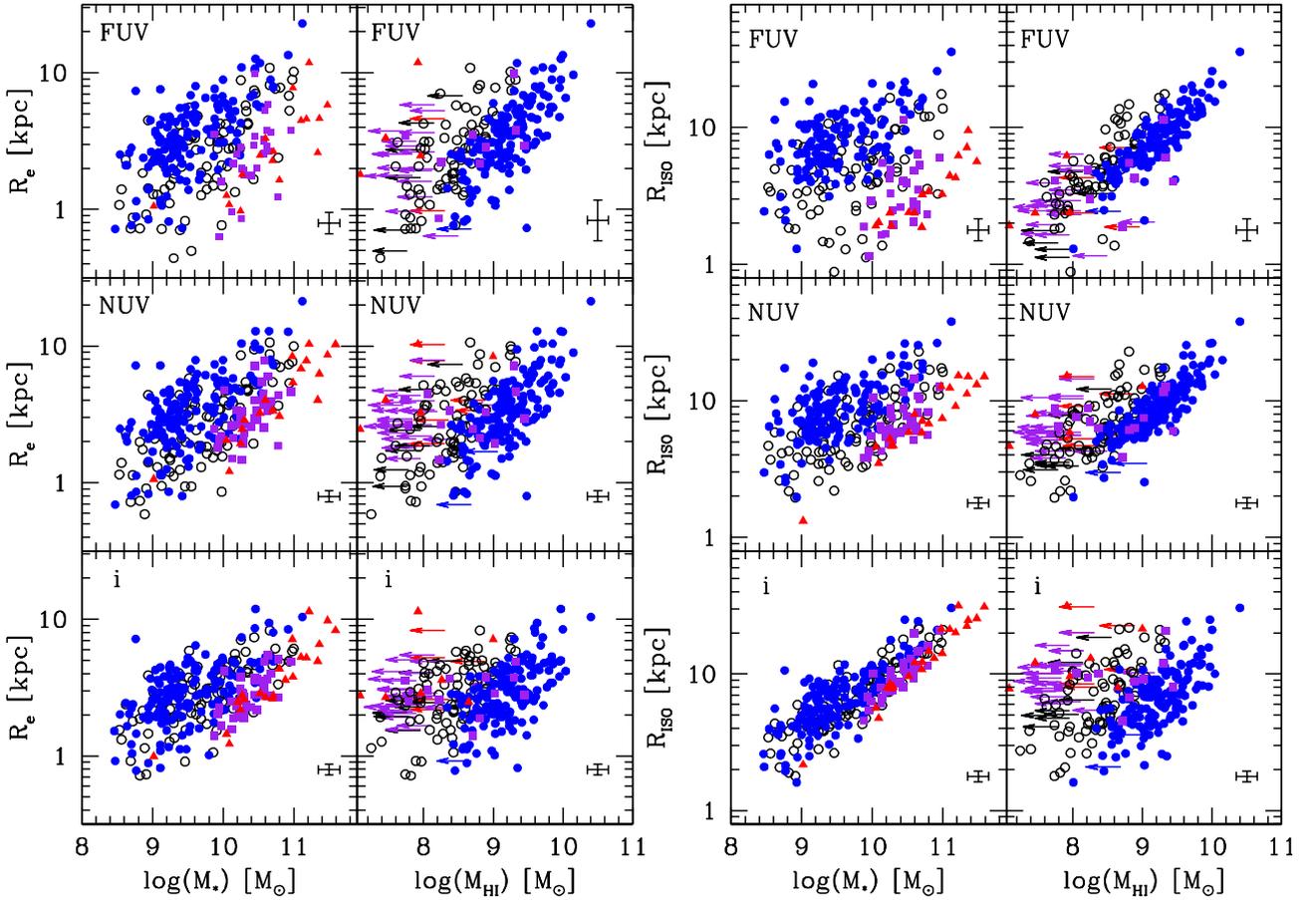}
     \caption{The radius vs. stellar mass and radius vs. \hi\ mass relations in FUV (top row), NUV (middle row) and $i$-band (bottom row). 
     The relations obtained for effective and isophotal radii are shown in the left and right panels, respectively. \hi-normal and \hi-deficient 
     spirals are indicated by filled and empty circles, respectively. Purple squares are S0 and S0a and red triangles E and dE.}
	 \label{radius}
  \end{figure*}

The goal of this exercise is just to establish whether or not this simple model is consistent with 
our interpretation, not to determine the best set of parameters fitting our observations. 
Since the model is based on pure disk galaxies, in Fig.~\ref{kormendylate} we compare the theoretical predictions 
with our observations for late-type galaxies only. We also separate \hi-normal (filled circles) from \hi-deficient 
galaxies (empty circles), in order to note any environmentally-driven 
trend in our analysis. The green and red tracks show the dust-free and reddened model, respectively. Spin parameters $\lambda$=0.02, 0.04 and 
0.06 are indicated by the dotted, solid and dashed line, respectively. 
In general, the model is able to reproduce the same trends observed in the data. In particular, the reversal in 
the $M_{*}$ vs. $<\mu_{e}>$ relation when moving from the optical to the UV is recovered. Moreover, as expected, although such change in slope 
is observed in both the dust-free and reddened model, the inclusion of the effects of dust makes it stronger. 
Finally, we note that the same scenario had been proposed by \cite{atlas2006} to explain the flattening of the UV surface brightness profiles 
in the inner parts of late-type spirals. We can thus conclude that the opposite trends observed 
in the optical and UV $M_{*}$ vs. $<\mu_{e}>$ relations are a natural consequence of the inside-out growth of stellar disks. 
It would be really interesting to investigate how this relation evolves with redshift, and at which epoch this reversal starts 
to appear. 

The same scenario invoked to explain the $M_{*}$ vs. $<\mu_{e}>$ relation is valid for the $<\mu_{e}>$ vs. $R_{e}$ relation 
(see Fig.~\ref{kormendylate}, top row).
Contrary to the previous case, here only a very weak correlation is observed in $i$-band, while in UV the surface brightness monotonically 
increases with effective radius. Moreover, in this case, early- and late-type galaxies show a similar trend although, at fixed radius, 
ellipticals and lenticulars are offset towards lower surface brightness than late-type systems. This is just a consequence 
of the different origin of the UV emission in the two morphological classes. Interestingly, the UV $<\mu_{e}>$ vs. $R_{e}$ relation 
is the one where \hi-deficient and \hi-normal galaxies show the largest difference, with \hi-poor systems having 
smaller radii and fainter $<\mu_{e}>$ than \hi-normal objects. As discussed in Sec.\ref{secenv}, this is consistent 
with a scenario in which the gas stripping is followed by the outside-in quenching of the star formation.

\section{Effective vs. Isophotal radii}
In the previous section we have focused our attention on the main scaling relations 
involving stellar mass and effective quantities. However, several studies have shown that isophotal radii often 
provide better (i.e., less scattered) scaling relations than effective ones. 
Some examples are the velocity vs. size relation, the stellar mass/luminosity vs. size relation \citep{saintonge11c} and 
the \hi\ mass vs. size relation, which is also usually adopted to calibrate the \hi\ deficiency parameter \citep{haynes,solanes96}. 
Thus, in Fig.\ref{radius}, we investigate whether isophotal radii provide less scattered relations than effective ones, by comparing 
the size vs. stellar mass and size vs. \hi\ mass relations in $i$, NUV and FUV bands. 
The properties of the best bisector linear fits are given in Table~\ref{risotab}.
 
Remarkably, when isophotal radii are used, the scatter in the size vs. \hi\ mass relation 
decreases significantly when moving from the optical to the UV, while the opposite trend is seen in the 
isophotal size vs. stellar mass relation. Thus, for the stellar mass vs. optical radius and \hi\ mass vs. UV radius relations, 
isophotal radii provide the smallest scatters. The dispersion in the $i$-band size 
vs. stellar mass relation decreases from $\sim$0.11 to $\sim$0.07 for E+S0a and from 0.20 to 0.13 dex for \hi-normal spirals, 
when effective radii are replaced by isophotal measurements. Similarly, the scatter in the UV size vs. \hi\ mass relation for 
\hi-normal late-type galaxies decreases from $\sim$0.17 dex to $\sim$0.1 dex in both NUV and FUV.

Our findings are easily explained by the fact that, contrary to effective radii, isophotal sizes are not affected by the presence of a central light concentration (such a bulge or a bar), and thus they represent a better proxy for the extent of the optical/UV disk. However, it is important to note that such a significant difference between 
isophotal and effective radii is only found when we consider two quantities that are expected to be strongly correlated by 
default: i.e., stellar mass vs. size of the stellar disk and \hi\ mass vs. size of the UV disk. 
Indeed, as shown in Table~\ref{scaletab} and \ref{risotab}, no strong difference in the scatter of the $i$-band size vs. \hi\ mass 
relation or the UV size vs. stellar mass relation for late-type galaxies is found when isophotal radii are used.

Looking at Fig. \ref{radius}, it is important to note the behavior of \hi-deficient galaxies in the 
radius vs. \hi\ mass relation. In $i$-band they are, by definition, systematically offset from the relation of \hi-normal galaxies, 
since the optical isophotal radius is indeed used to estimate the \hi-deficiency parameter. 
However, such offset gradually disappears when moving from optical to UV radii, suggesting that the UV is a very good 
tracer of atomic hydrogen \citep{donas87,bigiel10b,catinella10,cortese11}, regardless of the evolutionary history (e.g., secular or environmentally driven) of the galaxy. 

\begin{figure}
  \centering
  \includegraphics[width=9cm]{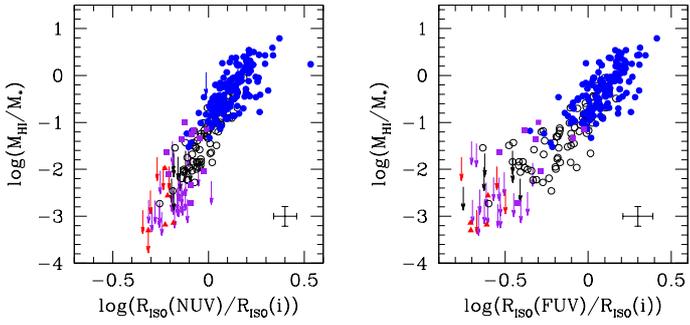}
     \caption{The \hi\ gas fraction vs. the NUV-to-$i$ (left) and FUV-to-$i$ (right) isophotal radii ratios. Symbols are as in Fig.~\ref{radius}. Arrows 
     indicates \hi\ non detections.}
	 \label{risogf}
  \end{figure}

The extremely tight optical $R_{ISO}$ vs. stellar mass and UV $R_{ISO}$ vs. \hi\ mass relations suggest that the 
extent of the star-forming (i.e., UV) disk normalized to the stellar mass (i.e., optical) one should be strongly correlated 
to the \hi\ gas fraction of a galaxy. Indeed, this is exactly what we find, as shown in Fig.~\ref{risogf}. 
Intriguingly, all galaxies detected in \hi\ seem to follow the same relation, supporting the idea that the UV and \hi\ emission 
from galaxies are tightly linked, in particular in the outer (dust-free) star-forming disk \citep{bigiel10b}, and have similar variations with galaxy properties and environment (see also next section).  
This result suggests that the amount of \hi\ (per unit of stellar mass) directly regulates the inside-out growth of the disk in spiral galaxies. 
This supports the recent results of \cite{wang11} who showed that, at fixed stellar mass, the higher the \hi\ gas fraction of a galaxy 
the bluer and more actively star-forming its outer disk is.

\section{The effect of the Virgo cluster on the UV morphology} 
\label{secenv}

\begin{figure}
  \centering
  \includegraphics[width=9cm]{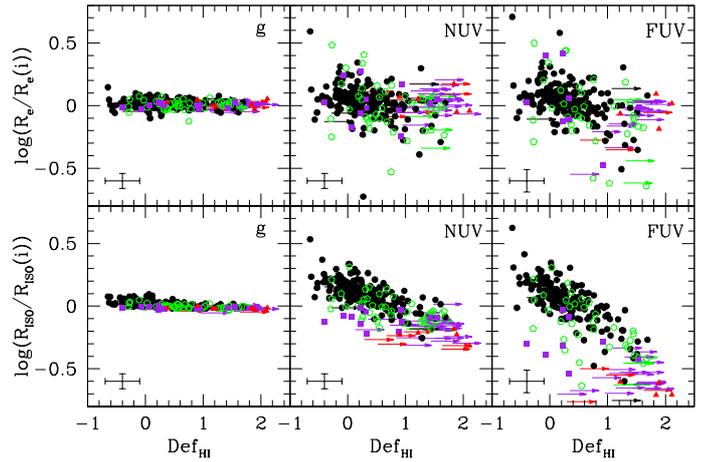}
     \caption{Top row: The $g$- (left), NUV- (middle), FUV-to-$i$ (right) effective radius ratio vs. \hi\ deficiency. 
     Bottom row:  Same for the isophotal radius ratio. Symbols are as in Fig.~\ref{kormendy}.}
	 \label{risodef}
  \end{figure}

The results presented in the previous section provide direct evidence of the strong connection between 
\hi\ content and extent of the UV star-forming disk. Since \hi\ removal is one of the most dramatic 
effects of the environment on cluster spirals \citep{cayatte90,ages1367,cortese11,chung09}, it is interesting to investigate what happens to 
the extent of the UV disk when the \hi\ is gradually removed from the galaxy. 
Fig~\ref{risogf} already suggests that the UV disk should shrink in \hi-poor galaxies, and this is 
clearly visible in Fig.~\ref{risodef}, where we plot the ratio of the $g$-to-$i$, NUV-to-$i$ and FUV-to-$i$ effective (top row) and 
isophotal radii  (bottom row) as a function of \hi-deficiency. The quantity in the y-axis is sometimes referred to 
as the {\it truncation parameter} (i.e., the ratio of the truncation radius to the radius of the old stellar population, \citealp{review}) 
and it has been used to investigate the effect of the environment on the \hi\ \citep{cayatte90,chung09}, 
H$\alpha$ \citep{review,rose10} and dust \citep{cortese10c} disks in \hi-deficient galaxies. This is the 
first time that such technique is applied to UV and optical data. 

If we focus on the ratio of isophotal radii, we clearly find that the extent of the NUV and FUV disks monotonically 
decreases with increasing \hi-deficiency ($r\sim -$0.6 and $-$0.7). However, this trend is strong only for late-type galaxies (i.e., Sa and later), whereas 
it becomes weaker for ellipticals and lenticulars ($r\sim -$0.35). This is not extremely surprising. Firstly, as already 
mentioned, in early type galaxies the UV emission may not trace young stellar populations, making it more difficult to justify a 
physical link between UV emission and \hi\ content. Secondly, the \hi\ deficiency is not well calibrated for early-type galaxies 
and it is not even clear whether a concept of \hi-deficiency is justified \citep{cortese11}. 
This suggests that the {\it truncation parameter} may not be a good indicator of environmental effects for early-type galaxies.

In the case of late-type galaxies, we find a clear change of slope in the relation between the {\it truncation parameter} and 
\hi-deficiency when moving from the FUV to the $g$-band. In particular, no trend is observed in the optical, suggesting that 
the truncation process has been quite recent and it has significantly affected only the UV morphology of the galaxy. 
This is entirely consistent with the predictions for a ram pressure stripping scenario \citep{n4569,dEale}.

By comparing the top and bottom rows of Fig.\ref{risodef}, it is clear that such a clean result is not obtained 
if effective radii are used to estimate the {\it truncation parameter}. Although a mild trend is still visible in the UV, it 
is more scattered and less significant. Indeed, the Pearson correlation coefficient for late-type galaxies decreases to $\sim -$0.3. 
This is just a natural consequence of the fact that the \hi\ stripping, the quenching of the star formation and the truncation of the 
UV disk start from the outskirts of a galaxy and gradually reach its inner parts. 
Thus, the effective radii are significantly less affected than the isophotal ones.

\begin{figure*}
  \centering
  \includegraphics[width=17cm]{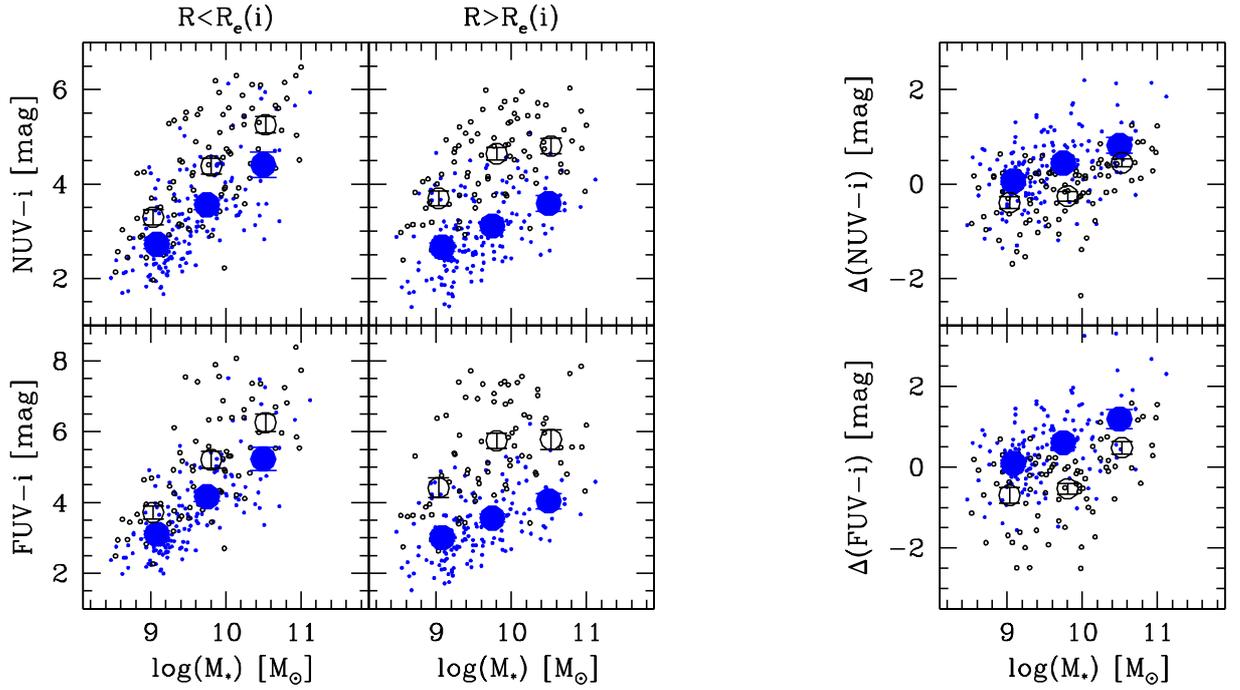}
     \caption{The inner (left) and outer (right) NUV-i (top) and FUV-i (bottom) colour vs. stellar mass. Symbols are as in Fig.~\ref{kormendylate}. 
     The right-most panel shows the inner-outer NUV-i (top) and FUV-i (bottom) colour difference as a function of stellar mass. The large circles 
     show the averages for each sub-sample in different bins of stellar mass.}
	 \label{gradient}
  \end{figure*}

In order to find additional support to this conclusion, we can investigate how much the colour (and thus the specific star formation rate, e.g., \citealp{schiminovich07}) 
in the inner regions of \hi-deficient galaxies are affected by the Virgo cluster environment. Thus, 
in Fig.~\ref{gradient} we compare the inner and outer $NUV-i$ (top panel) and $FUV-i$ (bottom panel) colour vs. stellar mass relation for 
\hi-normal and \hi-deficient spiral galaxies. Here, with inner and outer colours we refer to the colour inside and outside the 
$i$-band effective radius. We clearly find two different behaviors inside and outside the effective radius. 
In the inner parts, at fixed stellar mass, the colour of \hi-deficient galaxies is just $\sim$0.5-1 mag redder than 
the one observed in \hi-normal galaxies, and it follows a colour vs. stellar mass relation very similar to the one observed in Fig.~\ref{basic}. 
Conversely, in the outer parts, \hi-deficient systems are significantly redder ($\sim$1.5-2 mag) than \hi-normal galaxies and it is unclear 
if they still follow a colour vs. stellar mass relation. This automatically implies that, at fixed stellar mass, 
the shape of the colour gradients in \hi-deficient galaxies is altered (right panel in Fig.~\ref{gradient}) and, 
in the more extreme cases, its slope could even be inverted, i.e., showing inner parts bluer than the outer disk 
(see also \citealp{n4569}). 

This result does not only confirm that the inner parts of the UV disk are less affected 
during the stripping phase, but it also shows that no increase in the unobscured star formation activity follows the 
truncation of the star-forming disk. This is a very important result for our understanding of the effects of the environment 
on star formation. Indeed, previous studies of the H$\alpha$ morphology in nearby clusters have proposed an alternative scenario to 
ram pressure to explain the reduced extent of the star-forming disk in \hi-deficient galaxies (e.g., \citealp{moss98,moss00}). 
They suggested that the gravitational interaction could drive the flow of a significant fraction of gas into the central regions of a galaxy, 
triggering a starburst and thus altering the spatial distribution of the star-forming regions. Our findings seem to definitely rule 
out such a scenario, at least in environments similar to the Virgo cluster. Of course, our analysis cannot exclude 
the presence of a completely dust obscured starburst phase, and we will investigate this issue in a future paper. However, given 
the fact that ultra-luminous infrared galaxies are usually not observed in nearby clusters \citep{review}, 
we consider this possibility quite unlikely.

\section{Summary \& Conclusion}
In this paper we have presented UV and optical photometry and structural parameters for the HRS, a magnitude-, volume-limited 
sample of nearby galaxies in different environments. 
We used this new dataset to investigate, for the first time, the UV scaling relations and the effects of the environment 
on the UV morphology of nearby galaxies. 
Our results can be summarized as follows:
\begin{itemize}

\item We find a clear change of slope in the stellar mass vs. effective surface brightness relation when moving 
from the optical to the UV. In other words, massive galaxies have higher optical and lower UV surface brightnesses than 
less massive systems. By comparing our observations with the prediction of a simple multi-zone chemical model, we show 
that this is a direct consequence of the inside-out growth of the galactic disk combined with the fact that more massive 
systems have grown their disk earlier than dwarf galaxies. 

\item We show that isophotal radii almost always provide the tightest correlations with stellar and \hi\ masses than effective sizes. 
Particularly remarkable is the very low scatter in the correlation between UV isophotal radii and \hi\ mass, suggesting that the 
extent of the star-forming disk is directly linked to the amount of \hi\ in a galaxy. This conclusion is further confirmed by the 
fact that the ratio of UV-to-optical radius strongly correlates with the \hi\ gas fraction. 

\item We show that the tight connection between \hi\ content and size of the UV disk is also valid when environmentally perturbed \hi-deficient 
galaxies are included. In particular, we find a strong correlation between the size of the UV disk (normalized to the 
optical radius) and \hi-deficiency. Moreover, we show that, while the UV colour of the outer disk of 
\hi-deficient galaxies is significantly redder than that in \hi-normal galaxies, the galaxy center is less affected 
by the removal of the \hi. This is consistent with a simple truncation of the star-forming disk without any significant 
enhancement of the star formation in the inner parts. 

\end{itemize}

In conclusion, all our results are consistent with a steady inside-out growth of the UV disk in nearby galaxies via consumption 
of their \hi\ reservoir. Such growth can be stopped and even reversed only if the atomic hydrogen is removed via some kind of environmental 
effect.

\begin{acknowledgements}
We wish to thank the GALEX Team for their help and assistance during the planning of the observations 
and the GALEX TAC for allocating time to the HRS and GUViCS surveys. 
LC thanks Peppo Gavazzi for inspiring 
part of this work and Barbara Catinella, Peppo Gavazzi and Elysse Voyer for useful comments.
We thank the referee for useful suggestions that improved the clarity of this work.  

GALEX (Galaxy Evolution Explorer) is a NASA Small Explorer, launched in April 2003. 
We gratefully acknowledge NASA's support for construction, operation, and science analysis for the GALEX mission, 
developed in cooperation with the Centre National d'Etudes Spatiales (CNES) of France and the 
Korean Ministry of Science and Technology. 

Funding for the SDSS and SDSS-II has been provided by the Alfred P. Sloan Foundation, the Participating Institutions, 
the National Science Foundation, the U.S. Department of Energy, the National Aeronautics and Space Administration, 
the Japanese Monbukagakusho, the Max Planck Society, and the Higher Education Funding Council for England. 

The SDSS is managed by the Astrophysical Research Consortium for the Participating Institutions. 
The Participating Institutions are the American Museum of Natural History, Astrophysical Institute Potsdam, 
University of Basel, University of Cambridge, Case Western Reserve University, University of Chicago, Drexel University, 
Fermilab, the Institute for Advanced Study, the Japan Participation Group, Johns Hopkins University, the Joint Institute 
for Nuclear Astrophysics, the Kavli Institute for Particle Astrophysics and Cosmology, the Korean Scientist Group, 
the Chinese Academy of Sciences (LAMOST), Los Alamos National Laboratory, the Max-Planck-Institute for Astronomy (MPIA), 
the Max-Planck-Institute for Astrophysics (MPA), New Mexico State University, Ohio State University, 
University of Pittsburgh, University of Portsmouth, Princeton University, the United States Naval Observatory, 
and the University of Washington.

This research has made use of the NASA/IPAC Extragalactic Database (NED) which is operated 
by the Jet Propulsion Laboratory, California Institute of Technology, under contract 
with the National Aeronautics and Space Administration; and of the GOLDMine database \citep{goldmine}. 

The research leading to these results has received funding from the European Community's Seventh Framework Programme 
(/FP7/2007-2013/) under grant agreement No 229517.

\end{acknowledgements}

\begin{appendix}
\section{Notes on individual objects}
\begin{itemize}

\item{\bf HRS2.} Our photometry in the SDSS bands may be affected by the presence of 
a bright star $\sim$1 arcmin south-west from the target.

\item{\bf HRS3 \& HRS4.} Interacting system (Arp94). Photometry is uncertain due to 
the overlap between the two objects.

\item{\bf HRS20.} Interacting system (Arp270). Photometry is uncertain due to 
the overlap with the companion galaxy.

\item{\bf HRS42.} The point spread function (PSF) of the NUV image is significantly 
asymmetric and elongated towards North-West. However, this does not affect our integrated photometry 
and should not significantly influence the estimate of the structural parameters. 


\item{\bf HRS55.} The $i$-band frame could not be used due to the presence of large artifacts created 
by a nearby foreground bright star.

\item{\bf HRS60.} The $i$-band photometry is mildly affected by the presence of a satellite track.

\item{\bf HRS68.} The effective radius for this object is significantly smaller than the spatial 
resolution adopted in this analysis ($\sim$6 arcsec). Thus, no effective radius and surface brightness 
are provided. 

\item{\bf HRS74.} The PSF of the NUV image is significantly 
asymmetric and elongated towards North-East. However, this does not affect our integrated photometry 
and should not significantly influence the estimate of the structural parameters. 

\item{\bf HRS81.} The GALEX images suggest the presence of very low surface brightness loops/tidal features 
associated to this object, not clearly visible in SDSS. However, the data currently available are not deep 
enough to determine if these features are real. 

\item {\bf HRS105.} This galaxy has an extended UV ring, with ellipticity and position angle 
(0.5, +34 deg) significantly different from the ones determined from the $i$-band images 
(see also \citealp{cortese09,bettoni10}). However, the integrated magnitudes do not significantly 
change if these values are used for the profile fitting. 

\item {\bf HRS177.} The GALEX NUV image suggests the presence of a tail of star-forming knots departing 
from the galaxy towards the north (see also \citealp{arrigoni12}).

\item {\bf HRS202.} This galaxy has a typical FUV surface brightness fainter than 28 mag arcsec$^{-2}$, making
impossible to estimate its isophotal radius.

\item {\bf HRS211.} The galaxy center is saturated in the $i$-band, affecting our photometry.

\item {\bf HRS213.} This edge-on galaxy has a significant bulge component in the optical, which is 
not visible in the GALEX images. Thus the ellipticity adopted here is not a fair representation of 
the UV light distribution and could affect our estimate of the UV structural parameters.

\item {\bf HRS215 \& HRS216.}  Interacting system. Photometry is uncertain due to 
the overlap between the two objects.

\item {\bf HRS244 \& HRS245.}  Interacting system. Photometry is uncertain due to 
the overlap between the two objects.

\item{\bf HRS265.} The PSF of the NUV image is significantly 
asymmetric and elongated towards North-West. However, this does not affect our integrated photometry 
and should not significantly influence the estimate of the structural parameters.

\end{itemize}

\end{appendix}

\onecolumn
\begin{landscape}
\begin{center}
\scriptsize
\begin{longtable}{ccccccccccccccccc}
\caption[The general properties of the HRS sample and of the GALEX data used in the paper.]{The general properties of the HRS sample and of the GALEX data used in the paper.\label{tab:hrsgalaxies}}\\
\hline \hline \\[-2ex]
  \multicolumn{1}{c}{HRS} &
  \multicolumn{1}{c}{CGCG} &
  \multicolumn{1}{c}{VCC} &
  \multicolumn{1}{c}{UGC} &
  \multicolumn{1}{c}{NGC} &
  \multicolumn{1}{c}{IC} &
  \multicolumn{1}{c}{R.A.} &
  \multicolumn{1}{c}{Dec.} &
  \multicolumn{1}{c}{Type} &
  \multicolumn{1}{c}{$D_{25}$} &
  \multicolumn{1}{c}{E(B-V)} &
  \multicolumn{1}{c}{Dist.} &
  \multicolumn{1}{c}{Tile-Name NUV} &
  \multicolumn{1}{c}{t$_{exp}$ NUV} &
  \multicolumn{1}{c}{Tile-Name FUV} &
  \multicolumn{1}{c}{t$_{exp}$ FUV} &
  \multicolumn{1}{c}{NOTE} \\
  
    \multicolumn{1}{c}{} &
  \multicolumn{1}{c}{} &
  \multicolumn{1}{c}{} &
  \multicolumn{1}{c}{} &
  \multicolumn{1}{c}{} &
  \multicolumn{1}{c}{} &
  \multicolumn{1}{c}{(J.2000)} &
  \multicolumn{1}{c}{(J.2000)} &
  \multicolumn{1}{c}{} &
  \multicolumn{1}{c}{(arcmin)} &
  \multicolumn{1}{c}{(mag)} &
  \multicolumn{1}{c}{(Mpc)} &
  \multicolumn{1}{c}{} &
  \multicolumn{1}{c}{(sec)} &
  \multicolumn{1}{c}{} &
  \multicolumn{1}{c}{(sec)} &
  \multicolumn{1}{c}{} \\ [0.5ex] \hline
   \\[-1.8ex]
\endfirsthead
\multicolumn{17}{c}{{\tablename} \thetable{} -- Continued} \\[0.5ex]
  \hline \hline \\[-2ex]
  \multicolumn{1}{c}{HRS} &
  \multicolumn{1}{c}{CGCG} &
  \multicolumn{1}{c}{VCC} &
  \multicolumn{1}{c}{UGC} &
  \multicolumn{1}{c}{NGC} &
  \multicolumn{1}{c}{IC} &
  \multicolumn{1}{c}{R.A.} &
  \multicolumn{1}{c}{Dec.} &
  \multicolumn{1}{c}{Type} &
  \multicolumn{1}{c}{$D_{25}$} &
  \multicolumn{1}{c}{E(B-V)} &
  \multicolumn{1}{c}{Dist.} &
  \multicolumn{1}{c}{Tile-Name NUV} &
  \multicolumn{1}{c}{t$_{exp}$ NUV} &
  \multicolumn{1}{c}{Tile-Name FUV} &
  \multicolumn{1}{c}{t$_{exp}$ FUV} &
  \multicolumn{1}{c}{NOTE} \\
  
    \multicolumn{1}{c}{} &
  \multicolumn{1}{c}{} &
  \multicolumn{1}{c}{} &
  \multicolumn{1}{c}{} &
  \multicolumn{1}{c}{} &
  \multicolumn{1}{c}{} &
  \multicolumn{1}{c}{(J.2000)} &
  \multicolumn{1}{c}{(J.2000)} &
  \multicolumn{1}{c}{} &
  \multicolumn{1}{c}{(arcmin)} &
  \multicolumn{1}{c}{(mag)} &
  \multicolumn{1}{c}{(Mpc)} &
  \multicolumn{1}{c}{} &
  \multicolumn{1}{c}{(sec)} &
  \multicolumn{1}{c}{} &
  \multicolumn{1}{c}{(sec)} &
  \multicolumn{1}{c}{} \\ [0.5ex] \hline
   \\[-1.8ex]
\endhead 
 \\[-1.8ex] \hline
  \multicolumn{17}{c}{{Continued on next page\ldots}} \\
\endfoot

  \\[-1.8ex] \hline \hline
\endlastfoot
  1 & 123-035 & ... & ... & ... & ... & 10:17:39.66 & +22:48:35.9 & 13 & 1.0 & 0.0311 & 16.79 & GI6\_012001\_HRS1 & 1385.7 & ... & ... & N\\
  2 & 124-004 & ... & 5588 & ... & ... & 10:20:57.13 & +25:21:53.4 & 5 & 0.52 & 0.023 & 18.44 & GI6\_012002\_HRS2 & 1669.1 & AIS\_213 & 201.0 & ...\\
  3 & 94-026 & ... & 5617 & 3226 & ... & 10:23:27.01 & +19:53:54.7 & 0 & 3.16 & 0.0226 & 16.7 & ... & ... & ... & ... & S\\
  4 & 94-028 & ... & 5620 & 3227 & ... & 10:23:30.58 & +19:51:54.2 & 3 & 5.37 & 0.0226 & 16.4 & ... & ... & ... & ... & S\\
  5 & 94-052 & ... & ... & ... & 610 & 10:26:28.37 & +20:13:41.5 & 7 & 1.86 & 0.0203 & 16.71 & ... & ... & ... & ... & S\\
  6 & 154-016 & ... & 5662 & ... & ... & 10:27:01.16 & +28:38:21.9 & 5 & 3.31 & 0.0267 & 18.89 & GI3\_079013\_NGC3245 & 1622.15 & GI3\_079013\_NGC3245 & 1622.15 & ...\\
  7 & 154-017 & ... & 5663 & 3245 & ... & 10:27:18.39 & +28:30:26.6 & 1 & 3.24 & 0.0247 & 18.77 & GI3\_079013\_NGC3245 & 1622.15 & GI3\_079013\_NGC3245 & 1622.15 & ...\\
  8 & 154-020 & ... & 5685 & 3254 & ... & 10:29:19.92 & +29:29:29.2 & 6 & 5.01 & 0.0201 & 19.37 & GI6\_012003\_HRS8 & 1682.1 & AIS\_213 & 222.0 & ...\\
\end{longtable}
\end{center}                                                                                                      
{\scriptsize 
The columns are as follows: \\
Colums 1-6:  HRS \citep{HRS}, CGCG \citep{ZWHE61}, VCC \citep{vcc}, UGC \citep{ugc}, NGC \citep{ngc} and IC \citep{ic} names. \\
Column 7-8: the J.2000 right ascension and declination.\\
Column 9: Morphological type: -2=dE/dS0, 0=E-E/S0, 1=S0, 2=S0a-S0/Sa, 3=Sa, 4=Sab, 5=Sb, 6=Sbc, 7=Sc, 8=Scd, 
9=Sd, 10=Sdm-Sd/Sm, 11=Sm, 12=Im, 13=Pec, 14=S/BCD, 15=Sm/BCD, 16=Im/BCD, 17=BCD.\\
Column 10: Optical diameter ($D_{25}$) in arcminutes as given in \cite{HRS}. Note that for HRS295 we adopted the RC3 value instead of that 
given by \cite{HRS}, since it is much more representative of the `real' optical extent of the galaxy.\\
Column 11: $E(B-V)$ based on the maps of \cite{schlegel98}.\\
Column 12: Distance in Mpc. As discussed in \cite{HRS}, we fixed the distances for galaxies belonging to the Virgo cluster (i.e., 23 Mpc for the 
Virgo B cloud and 17 Mpc for all the other clouds; \citealp{gav99}), while for the rest of the sample 
distances have been estimated from their recessional velocities assuming a Hubble constant $H_{0}=$70 km s$^{-1}$ Mpc$^{-1}$.\\
Column 13-14: Name of the NUV tile and exposure time in seconds.\\
Column 15-16: Name of the FUV tile and exposure time in seconds.\\
Column 17: Note indicating whether a galaxy has been excluded because GALEX 
observations were not possible due to bright stars (S), the target was too close to 
the edge of the field (E), the frame was too shallow to be used for reliable 
photometry (F) or simply the galaxy was not observed (N). 
}
\end{landscape}

\twocolumn
\onecolumn
\begin{landscape}
\begin{center}
\scriptsize
\begin{longtable}{ccccccccccccccccccccccccc}
\caption[Integrated FUV, NUV, $g$, $r$ and $i$ magnitudes and stellar masses for the HRS.]{Integrated FUV, NUV, $g$, $r$ and $i$ magnitudes and stellar masses for the HRS.\label{tab:hrsmags}}\\
\hline \hline \\[-2ex]
  \multicolumn{1}{c}{HRS} &
  \multicolumn{1}{c}{e} &
  \multicolumn{1}{c}{PA} &
  \multicolumn{2}{c}{FUV$_{asy}$} &
  \multicolumn{2}{c}{FUV$_{D_{25}}$} &
  \multicolumn{2}{c}{NUV$_{asy}$} &
  \multicolumn{2}{c}{NUV$_{D_{25}}$} &
  \multicolumn{2}{c}{$g_{asy}$} &
  \multicolumn{2}{c}{$g_{D_{25}}$} &
  \multicolumn{2}{c}{$r_{asy}$} &
  \multicolumn{2}{c}{$r_{D_{25}}$} &
  \multicolumn{2}{c}{$i_{asy}$} &
  \multicolumn{2}{c}{$i_{D_{25}}$} &
  \multicolumn{1}{c}{$\log(M_{*})$} \\
  
  \multicolumn{1}{c}{} &
  \multicolumn{1}{c}{} &
  \multicolumn{1}{c}{($\deg$)} &
  \multicolumn{2}{c}{(mag)} &
  \multicolumn{2}{c}{(mag)} &
  \multicolumn{2}{c}{(mag)} &
  \multicolumn{2}{c}{(mag)} &
  \multicolumn{2}{c}{(mag)} &
  \multicolumn{2}{c}{(mag)} &
  \multicolumn{2}{c}{(mag)} &
  \multicolumn{2}{c}{(mag)} &
  \multicolumn{2}{c}{(mag)} &
  \multicolumn{2}{c}{(mag)} & 
  \multicolumn{1}{c}{(M$_{\odot}$)}\\ [0.5ex] \hline
   \\[-1.8ex]
\endfirsthead
\multicolumn{24}{c}{{\tablename} \thetable{} -- Continued} \\[0.5ex]
  \hline \hline \\[-2ex]
  \multicolumn{1}{c}{HRS} &
  \multicolumn{1}{c}{e} &
  \multicolumn{1}{c}{PA} &
  \multicolumn{2}{c}{FUV$_{asy}$} &
  \multicolumn{2}{c}{FUV$_{D_{25}}$} &
  \multicolumn{2}{c}{NUV$_{asy}$} &
  \multicolumn{2}{c}{NUV$_{D_{25}}$} &
  \multicolumn{2}{c}{$g_{asy}$} &
  \multicolumn{2}{c}{$g_{D_{25}}$} &
  \multicolumn{2}{c}{$r_{asy}$} &
  \multicolumn{2}{c}{$r_{D_{25}}$} &
  \multicolumn{2}{c}{$i_{asy}$} &
  \multicolumn{2}{c}{$i_{D_{25}}$} &
  \multicolumn{1}{c}{$\log(M_{*})$} \\
  
  \multicolumn{1}{c}{} &
  \multicolumn{1}{c}{} &
  \multicolumn{1}{c}{($\deg$)} &
  \multicolumn{2}{c}{(mag)} &
  \multicolumn{2}{c}{(mag)} &
  \multicolumn{2}{c}{(mag)} &
  \multicolumn{2}{c}{(mag)} &
  \multicolumn{2}{c}{(mag)} &
  \multicolumn{2}{c}{(mag)} &
  \multicolumn{2}{c}{(mag)} &
  \multicolumn{2}{c}{(mag)} &
  \multicolumn{2}{c}{(mag)} &
  \multicolumn{2}{c}{(mag)} & 
  \multicolumn{1}{c}{(M$_{\odot}$)}\\ [0.5ex] \hline
   \\[-1.8ex]
\endhead 
 \\[-1.8ex] \hline
  \multicolumn{24}{c}{{Continued on next page\ldots}} \\
\endfoot

  \\[-1.8ex] \hline \hline
\endlastfoot

         1 &       0.60 &         -5 &      ...   &      ...   &       ...   &      ...   &      17.680 &      0.031 &      17.759 &      0.034 &      14.870 &      0.030 &      14.944 &      0.032 &      14.233 &      0.021 &      14.307 &      0.023 &      13.899 &      0.031 &      13.995 &      0.032  &        8.72 \\
         2 &       0.28 &         35 &     16.322 &      0.059 &      16.488 &      0.092 &      15.878 &      0.030 &      16.034 &      0.071 &      13.951 &      0.030 &      14.083 &      0.066 &      13.480 &      0.020 &      13.612 &      0.060 &      13.259 &      0.030 &      13.407 &      0.064  &        8.77 \\
         3 &       0.17 &         17 &      ...   &      ...   &       ...   &      ...   &      ...    &      ...   &       ...   &      ...   &      11.821 &      0.036 &      11.975 &      0.036 &      11.078 &      0.022 &      11.220 &      0.022 &      10.661 &      0.031 &      10.809 &      0.032  &       10.21 \\
         4 &       0.48 &        -33 &      ...   &      ...   &       ...   &      ...   &      ...    &      ...   &       ...   &      ...   &      11.189 &      0.034 &      11.256 &      0.036 &      10.552 &      0.021 &      10.599 &      0.022 &      10.162 &      0.031 &      10.214 &      0.031  &       10.26 \\
         5 &       0.76 &         29 &      ...   &      ...   &       ...   &      ...   &      ...    &      ...   &       ...   &      ...   &      13.920 &      0.034 &      14.040 &      0.033 &      13.057 &      0.023 &      13.164 &      0.022 &      12.590 &      0.032 &      12.674 &      0.032  &        9.62 \\
         6 &       0.85 &        -30 &     17.499 &      0.052 &      17.628 &      0.055 &      17.005 &      0.032 &      17.114 &      0.034 &      14.489 &      0.032 &      14.632 &      0.034 &      13.923 &      0.029 &      14.046 &      0.026 &      13.545 &      0.038 &      13.742 &      0.042  &        8.94 \\
\end{longtable}
\end{center}
{\scriptsize 
The columns are as follows:\\
Column 1: HRS name, repeated from Table ~\ref{tab:hrsgalaxies}.\\
Column 2-3: Ellipticity and position angle (in degrees, measured counterclockwise from the North) of the ellipses used for the photometry. \\
Column 4-5: Asymptotic FUV magnitude and error. \\
Column 6-7: FUV magnitude measured within the optical diameter and error.\\
Column 8-9: Asymptotic NUV magnitude and error. \\
Column 10-11: NUV magnitude measured within the optical diameter and error.\\
Column 12-13: Asymptotic $g$ magnitude and error. \\
Column 14-15: $g$ magnitude measured within the optical diameter and error.\\
Column 16-17: Asymptotic $r$ magnitude and error. \\
Column 18-19: $r$ magnitude measured within the optical diameter and error.\\
Column 20-21: Asymptotic $i$ magnitude and error. \\
Column 22-23: $i$ magnitude measured within the optical diameter and error.\\
Column 24: Stellar masses determined from $i$-band luminosities $L_{i}$ using the $g-i$ colour-dependent 
stellar mass-to-light ratio relation from \cite{zibetti09}, assuming a \cite{chabrier} initial mass function (IMF):
$\log(M_{*}/L_{i}) = -0.963 + 1.032*(g-i)$. The $i$ luminosity and $g-i$ colour have been corrected  for Galactic extinction 
assuming a \cite{cardelli89} extinction law with $A(V)/E(B-V)=$3.1: i.e., $A(\lambda)/E(B-V)$=3.793 and 2.086 for $g$ 
and $i$, respectively. The typical uncertainty in this estimate is $\sim$0.15 dex.
}
\end{landscape}

\twocolumn
\onecolumn
\begin{landscape}
\begin{center}
\scriptsize
\begin{longtable}{cccccccccccccccc}
\caption[FUV, NUV, $g$, $r$ and $i$ effective and isophotal radii and effective surface brigthnesses for the HRS.]{FUV, NUV, $g$, $r$ and $i$ effective and isophotal radii and effective surface brigthnesses for the HRS.\label{tab:struct}}\\
\hline \hline \\[-2ex]
  \multicolumn{1}{c}{HRS} &
  \multicolumn{1}{c}{R$_{e}$(FUV)} &
  \multicolumn{1}{c}{R$_{ISO}$(FUV)} &
  \multicolumn{1}{c}{$<\mu_{e}>$ (FUV)} &
  \multicolumn{1}{c}{R$_{e}$(NUV)} &
  \multicolumn{1}{c}{R$_{ISO}$(NUV)} &
  \multicolumn{1}{c}{$<\mu_{e}>$ (NUV)} &
  \multicolumn{1}{c}{R$_{e}$($g$)} &
  \multicolumn{1}{c}{R$_{ISO}$($g$)} &
  \multicolumn{1}{c}{$<\mu_{e}>$ ($g$)} &
  \multicolumn{1}{c}{R$_{e}$($r$)} &
  \multicolumn{1}{c}{R$_{ISO}$($r$)} &
  \multicolumn{1}{c}{$<\mu_{e}>$ ($r$)} &
  \multicolumn{1}{c}{R$_{e}$($i$)} &
  \multicolumn{1}{c}{R$_{ISO}$($i$)} &
  \multicolumn{1}{c}{$<\mu_{e}>$ ($i$)} \\
  
  \multicolumn{1}{c}{} &
  \multicolumn{1}{c}{(arcsec)} &
  \multicolumn{1}{c}{(arcsec)} &
  \multicolumn{1}{c}{(mag arcsec$^{-2}$)} &
  \multicolumn{1}{c}{(arcsec)} &
  \multicolumn{1}{c}{(arcsec)} &
  \multicolumn{1}{c}{(mag arcsec$^{-2}$)} &
  \multicolumn{1}{c}{(arcsec)} &
  \multicolumn{1}{c}{(arcsec)} &
  \multicolumn{1}{c}{(mag arcsec$^{-2}$)} &
  \multicolumn{1}{c}{(arcsec)} &
  \multicolumn{1}{c}{(arcsec)} &
  \multicolumn{1}{c}{(mag arcsec$^{-2}$)} &
  \multicolumn{1}{c}{(arcsec)} &
  \multicolumn{1}{c}{(arcsec)} &
  \multicolumn{1}{c}{(mag arcsec$^{-2}$)} \\ [0.5ex] \hline
   \\[-1.8ex]
\endfirsthead

\multicolumn{16}{c}{{\tablename} \thetable{} -- Continued} \\[0.5ex]
\hline \hline \\[-2ex]
  \multicolumn{1}{c}{HRS} &
  \multicolumn{1}{c}{R$_{e}$(FUV)} &
  \multicolumn{1}{c}{R$_{ISO}$(FUV)} &
  \multicolumn{1}{c}{$<\mu_{e}>$ (FUV)} &
  \multicolumn{1}{c}{R$_{e}$(NUV)} &
  \multicolumn{1}{c}{R$_{ISO}$(NUV)} &
  \multicolumn{1}{c}{$<\mu_{e}>$ (NUV)} &
  \multicolumn{1}{c}{R$_{e}$($g$)} &
  \multicolumn{1}{c}{R$_{ISO}$($g$)} &
  \multicolumn{1}{c}{$<\mu_{e}>$ ($g$)} &
  \multicolumn{1}{c}{R$_{e}$($r$)} &
  \multicolumn{1}{c}{R$_{ISO}$($r$)} &
  \multicolumn{1}{c}{$<\mu_{e}>$ ($r$)} &
  \multicolumn{1}{c}{R$_{e}$($i$)} &
  \multicolumn{1}{c}{R$_{ISO}$($i$)} &
  \multicolumn{1}{c}{$<\mu_{e}>$ ($i$)} \\
  
  \multicolumn{1}{c}{} &
  \multicolumn{1}{c}{(arcsec)} &
  \multicolumn{1}{c}{(arcsec)} &
  \multicolumn{1}{c}{(mag arcsec$^{-2}$)} &
  \multicolumn{1}{c}{(arcsec)} &
  \multicolumn{1}{c}{(arcsec)} &
  \multicolumn{1}{c}{(mag arcsec$^{-2}$)} &
  \multicolumn{1}{c}{(arcsec)} &
  \multicolumn{1}{c}{(arcsec)} &
  \multicolumn{1}{c}{(mag arcsec$^{-2}$)} &
  \multicolumn{1}{c}{(arcsec)} &
  \multicolumn{1}{c}{(arcsec)} &
  \multicolumn{1}{c}{(mag arcsec$^{-2}$)} &
  \multicolumn{1}{c}{(arcsec)} &
  \multicolumn{1}{c}{(arcsec)} &
  \multicolumn{1}{c}{(mag arcsec$^{-2}$)} \\ [0.5ex] \hline
   \\[-1.8ex]
\endhead 
 \\[-1.8ex] \hline
  \multicolumn{16}{c}{{Continued on next page\ldots}} \\
\endfoot

  \\[-1.8ex] \hline \hline
\endlastfoot

         1  &         ...  &         ...  &        ...   &        11.5  &        31.3  &       23.99  &       11.2  &        27.6  &       21.11  &        11.0  &        28.4 &       20.43  &        11.0  &        27.4  &       20.11 \\
         2  &         9.9  &        27.8  &       22.94  &         9.6  &        30.5  &       22.43  &        8.9  &        22.0  &       20.34  &         8.8  &        22.0 &       19.83  &         8.7  &        21.7  &       19.60 \\
         3  &         ...  &         ...  &        ...   &         ...  &         ...  &        ...   &       32.7  &        94.4  &       21.19  &        30.8  &       100.6 &       20.32  &        30.7  &        98.0  &       19.89 \\
         4  &         ...  &         ...  &        ...   &         ...  &         ...  &        ...   &       51.2  &       143.9  &       21.02  &        49.7  &       143.6 &       20.32  &        49.3  &       140.1  &       19.91 \\
         5  &         ...  &         ...  &        ...   &         ...  &         ...  &        ...   &       28.2  &        58.9  &       21.62  &        26.3  &        60.5 &       20.60  &        24.9  &        60.2  &       20.02 \\
         6  &        49.2  &        81.8  &       25.89  &        47.9  &        99.1  &       25.34  &       41.1  &        66.3  &       22.49  &        39.4  &        68.9 &       21.83  &        42.6  &        65.0  &       21.63 \\
         7  &        21.2  &        27.0  &       26.28  &        29.1  &        64.6  &       25.11  &       30.0  &        93.8  &       20.28  &        29.1  &        95.2 &       19.56  &        29.6  &        93.0  &       19.20 \\
         8  &        81.1  &       142.9  &       25.25  &        78.5  &       143.5  &       24.87  &       55.0  &       117.6  &       21.67  &        48.2  &       115.4 &       20.81  &        45.1  &       110.8  &       20.33 \\
\end{longtable}
\end{center}
{\scriptsize
The columns are as follows:\\
Column 1: HRS name, repeated from Table ~\ref{tab:hrsgalaxies}.\\
Column 2-4: Effective radius, isophotal radius at 28 mag arcsec$^{-2}$ level and effective surface brightness in FUV. \\
Column 5-7: Effective radius, isophotal radius at 28 mag arcsec$^{-2}$ level and effective surface brightness in NUV. \\
Column 8-10: Effective radius, isophotal radius at 24.5 mag arcsec$^{-2}$ level and effective surface brightness in $g$. \\
Column 11-13: Effective radius, isophotal radius at 24. mag arcsec$^{-2}$ level and effective surface brightness in $r$.\\ 
Column 14-16: Effective radius, isophotal radius at 23.5 mag arcsec$^{-2}$ level and effective surface brightness in $i$. 
}
\end{landscape}

\begin{table}
{\scriptsize
\begin{center}
\caption {Best bysector linear fitting, Pearson correlation coefficients and dispersion perpendicolar to the best-fit for the structural scaling relations presented in \S~5. 
The linear fitting parameters are provided only when the Pearson correlation coefficient gives a probability higher than $\sim$95\% that the two variables are correlated.}
\label{scaletab}
\begin{tabular}{ccccccc}
\hline\hline
\noalign{\smallskip}
Sample  & Band &  N   &a  &  b  &  r  & $\sigma$\\
\noalign{\smallskip}
\hline
\noalign{\smallskip}
                                    & 	    &    \multicolumn{5}{c}{$\log(R_{50})=a\times\log(M_{*})+b$} \\
\noalign{\smallskip}
  All     	                    &  $i$  & 	311 & 0.38$\pm$0.02    & -3.29$\pm$0.09  & 0.58   &   0.21  \\	       
	  	                    &  NUV  & 	298 & 0.45$\pm$0.03    & -3.95$\pm$0.10  & 0.60   &   0.24  \\	       
	  	                    &  FUV  & 	265 & 0.53$\pm$0.05    & -4.74$\pm$0.14  & 0.48   &   0.30  \\	       
\noalign{\smallskip}
  E-S0a   	                    &  $i$  & 	61  &  0.48$\pm$0.03	& -4.50$\pm$0.18  & 0.84   &  0.10   \\	       
	  	                    &  NUV  & 	58  &  0.51$\pm$0.04	& -4.88$\pm$0.24  & 0.78   &  0.14    \\	       
	  	                    &  FUV  & 	44  &  0.64$\pm$0.09	& -6.36$\pm$0.44  & 0.62   &  0.21    \\	       
\noalign{\smallskip}
  Sa-Sab  	                    &  $i$  & 	48  & 0.43$\pm$0.03    & -3.87$\pm$0.19  & 0.81   &   0.14   \\	       
	  	                    &  NUV  & 	47  & 0.60$\pm$0.04    & -5.57$\pm$0.25  & 0.84   &   0.16   \\	       
	  	                    &  FUV  & 	40  & 0.69$\pm$0.06    & -6.43$\pm$0.35  & 0.80   &   0.20   \\	       
\noalign{\smallskip}
  $\geq$Sb                          &  $i$  &  202  & 0.44$\pm$0.03	& -3.78$\pm$0.12  & 0.61   &  0.20   \\    
	          	     	    &  NUV  &  193  &  0.50$\pm$0.03	& -4.32$\pm$0.14  & 0.67   &  0.20   \\	    
	          	     	    &  FUV  &  181  &  0.51$\pm$0.03	& -4.39$\pm$0.15  & 0.65   &  0.21   \\	    
\noalign{\smallskip}
  $\geq$Sa \& $Def_{HI}<$0.5        &  $i$  &  155  & 0.44$\pm$0.04    & -3.77$\pm$0.14  & 0.60   &    0.20   \\	       
	  		     	    &  NUV  &  149  &  0.52$\pm$0.04	& -4.49$\pm$0.16  & 0.65   &   0.20   \\	     
	  		     	    &  FUV  &  141  &  0.53$\pm$0.04	& -4.56$\pm$0.17  & 0.67   &   0.20   \\	     
\noalign{\smallskip}
  $\geq$ Sa~ \&~ $Def_{HI}\geq$0.5  &  $i$  & 	89  & 0.39$\pm$0.03    & -3.36$\pm$0.15  & 0.69   &    0.18   \\	       
	    			    &  NUV  & 	86  &  0.49$\pm$0.04	& -4.31$\pm$0.18  & 0.73   &   0.20   \\	 
	    			    &  FUV  & 	78  &  0.54$\pm$0.05	& -4.88$\pm$0.24  & 0.67   &   0.24   \\	 

\noalign{\smallskip}							    	  								  
\hline
\noalign{\smallskip}							    	  								  
                                    & 	    &    \multicolumn{5}{c}{$<\mu_{e}>=a\times\log(M_{*})+b$} \\
\noalign{\smallskip}
  All	 			    &  $i$  &  311  & -1.22$\pm$0.06  & +32.0$\pm$0.6  & $-$0.61   &   0.46   \\
	 			    &  NUV  &  298  & 1.61$\pm$0.11  & +8.46$\pm$0.90  & 0.51   &   0.59      \\
	 			    &  FUV  &  265  & 1.96$\pm$0.13  & +5.63$\pm$1.01  & 0.56   &   0.58      \\
\noalign{\smallskip}
  E-S0a 			     &  $i$  &  61   &  ...   & ...  & $-$0.13   &  ...      \\
				     &  NUV  &  58   &  ...   & ...  & 0.13      &  ...      \\      
				     &  FUV  &  44   &  ...   & ...  & $-$0.04   &  ...       \\
\noalign{\smallskip}
  Sa-Sab			     &  $i$  &  48   &  -1.00$\pm$0.38  & +29.8$\pm$2.4  & $-$0.30   &  0.46       \\
				     &  NUV  &  47   &  2.51$\pm$0.42  & -0.51$\pm$3.08  & 0.53   &	0.55    \\
				     &  FUV  &  40   &  3.28$\pm$0.52  & -7.51$\pm$3.81  & 0.58   &	0.52    \\
\noalign{\smallskip}
  $\geq$ Sb			     &  $i$  &  202  & -1.22$\pm$0.12  & +32.1$\pm$1.0  & $-$0.49   &	 0.46     \\
				     &  NUV  &  193  &   1.43$\pm$0.20  & +10.27$\pm$1.25  & 0.36   &	 0.59    \\
				     &  FUV  &  181  &  1.69$\pm$0.16  & +8.29$\pm$1.19  & 0.50   &	 0.52   \\
\noalign{\smallskip}
 $\geq$Sa \& $Def_{HI}<$0.5	     &  $i$  &  155  & -1.23$\pm$0.15  & +32.2$\pm$1.1  & $-$0.44   &  0.47	      \\
				     &  NUV  &  149  &   1.63$\pm$0.23  & +8.19$\pm$1.48  & 0.39   &	  0.56   \\
				     &  FUV  &  141  &   1.85$\pm$0.19  & +6.61$\pm$1.43  & 0.50   &	0.51     \\
\noalign{\smallskip}
  $\geq$Sa \& $Def_{HI}\geq$0.5      &  $i$   &  89  & -1.12$\pm$0.11  & +31.2$\pm$1.2  & $-$0.61   &   0.41      \\
				     &  NUV  &   86  &   1.56$\pm$0.26  & +9.18$\pm$1.85  & 0.43   &	 0.62    \\
				     &  FUV  &   78  &   2.00$\pm$0.30  & +5.53$\pm$2.14  & 0.48   &	 0.63    \\
\noalign{\smallskip}							    	  								  
\hline
\noalign{\smallskip}							    	  								  
                                    & 	    &    \multicolumn{5}{c}{$<\mu_{e}>=a\times\log(R_{50})+b$} \\
\noalign{\smallskip}
  All				       &  $i$  &  311  & 1.73$\pm$0.40  & +19.3$\pm$0.4 & 0.18   &  0.43      \\
				       &  NUV  &  308  & 3.64$\pm$0.24  & +22.4$\pm$0.5  & 0.53   &  0.29	    \\
				       &  FUV  &  272  & 3.75$\pm$0.31  & +23.0$\pm$0.6  & 0.47   & 0.35	    \\
\noalign{\smallskip}
  E-S0a 				&  $i$ & 61    &  ...              &  ...             & 0.27  &   ...      \\
				       &  NUV  & 59    &  3.01$\pm$0.45  & +23.6$\pm$1.3  & 0.53   &	0.23     \\
				       &  FUV  & 45    & 3.89$\pm$0.60  & +24.4$\pm$1.5  & 0.58   &   0.26	    \\
\noalign{\smallskip}
  Sa-Sab			       &  $i$  &  48   &  ...             & ...	         & 0.18    &	...     \\
				       &  NUV  &  48  & 4.52$\pm$0.49  & +22.5$\pm$1.3  & 0.70   &   0.26	    \\
				       &  FUV  &  40   & 5.15$\pm$0.57  & +23.0$\pm$1.4  & 0.72   &  0.29	    \\
\noalign{\smallskip}
  $\geq$ Sb			       &  $i$  &  202  & 1.96$\pm$0.35  & +19.6$\pm$0.5        & 0.28	 & 0.34	   \\
				       &  NUV  &  201     & 3.10$\pm$0.26  & +22.4$\pm$0.7  & 0.52   &	0.28    \\
				       &  FUV  &  187   &  3.33$\pm$0.28  & +22.7$\pm$0.7  & 0.52   &	0.29     \\
\noalign{\smallskip}
  $\geq$Sa \& $Def_{HI}<$0.5	       &  $i$  &  155  & 2.23$\pm$0.36  & +19.4$\pm$0.6  & 0.34	& 0.30	  \\
				       &  NUV  &  156  & 3.73$\pm$0.24  & +21.8$\pm$0.7  & 0.66   &	0.22    \\
				       &  FUV  &  147  & 3.91$\pm$0.24  & +22.1$\pm$0.7  & 0.69   &	0.22    \\
\noalign{\smallskip}
  $\geq$Sa \& $Def_{HI}\geq$0.5        &  $i$  &   89  & ...              & ...              & 0.05   &    ...     \\
				       &  NUV  &   87  & 3.49$\pm$0.35  & +22.9$\pm$1.0  & 0.61   &  0.26   \\
				       &  FUV  &   78    & 4.06$\pm$0.43  & +23.4$\pm$1.1  & 0.61   &	0.30    \\

\noalign{\smallskip}							     								 
\noalign{\smallskip}					        
\hline
\hline
\end{tabular}
\end{center}
}
\end{table}

\onecolumn
\begin{table}
{\scriptsize
\begin{center}
\caption {Best bysector linear fitting, Pearson correlation coefficients and dispersion perpendicolar to the best-fit for the radius vs. stellar mass 
and radius vs. \hi\ mass presented in Fig.~\ref{radius}. 
The best linear fitting parameters are provided only when the Pearson correlation coefficient gives a probability higher than $\sim$95\% that the two variables are correlated.}
\label{risotab}
\begin{tabular}{ccccccc}
\hline\hline
\noalign{\smallskip}
Sample  & Band &  N   &a  &  b  &  r  & $\sigma$\\
\noalign{\smallskip}
\hline
\noalign{\smallskip}
                                    & 	    &    \multicolumn{5}{c}{$\log(R_{ISO})=a\times\log(M_{*})+b$} \\
\noalign{\smallskip}
  All     	                    &  $i$  & 	312 & 0.38$\pm$0.01    & -2.83$\pm$0.06  & 0.85   &   0.13  \\	       
	  	                    &  NUV  & 	299 & 0.42$\pm$0.04    & -3.20$\pm$0.10  & 0.47   &   0.24  \\	       
	  	                    &  FUV  & 	265 & ...              & ...             & 0.11   &   ...     \\	       
\noalign{\smallskip}
  E-S0a   	                    &  $i$  & 	61  &  0.48$\pm$0.02	& -4.04$\pm$0.11  & 0.94   &  0.07   \\	       
	  	                    &  NUV  & 	58  &  0.43$\pm$0.03	& -3.69$\pm$0.19  & 0.80   &  0.11    \\	       
	  	                    &  FUV  & 	43  &  0.59$\pm$0.10	& -5.75$\pm$0.42  & 0.59   &  0.18    \\	       
\noalign{\smallskip}
  $\geq$Sa \& $Def_{HI}<$0.5        &  $i$  &  156  & 0.41$\pm$0.02    & -3.12$\pm$0.10  & 0.80   &    0.13   \\	       
	  		     	    &  NUV  &  150  & 0.42$\pm$0.03    & -3.04$\pm$0.13  & 0.63   &   0.17   \\	     
	  		     	    &  FUV  &  142  &  0.47$\pm$0.05	& -3.55$\pm$0.16  & 0.54   &   0.21   \\	     
\noalign{\smallskip}
  $\geq$ Sa~ \&~ $Def_{HI}\geq$0.5  &  $i$  & 	89  & 0.41$\pm$0.02    & -3.15$\pm$0.11  & 0.87   &    0.12   \\	       
	    			    &  NUV  & 	86  & 0.40$\pm$0.04	& -3.11$\pm$0.16  & 0.68   &   0.18   \\	 
	    			    &  FUV  & 	78  & 0.55$\pm$0.10	& -4.68$\pm$0.28  & 0.45   &   0.30   \\	 

\noalign{\smallskip}							    	  								  
\hline
\noalign{\smallskip}							    	  								  
                                    & 	    &    \multicolumn{5}{c}{$\log(R_{ISO})=a\times\log(M_{HI})+b$} \\
\noalign{\smallskip}
 $\geq$Sa \& $Def_{HI}<$0.5	     &  $i$  &  156  &   0.56$\pm$0.03  &  -4.3$\pm$0.2  & 0.69    &  0.16	      \\
				     &  NUV  &  157  &   0.50$\pm$0.02  &  -3.6$\pm$0.1  & 0.85    &  0.10      \\
				     &  FUV  &  148  &   0.53$\pm$0.02  &  -4.0$\pm$0.1  & 0.87    &  0.10     \\
\noalign{\smallskip}							    	  								  
\hline
\noalign{\smallskip}							    	  								  
                                    & 	    &    \multicolumn{5}{c}{$\log(R_{50})=a\times\log(M_{HI})+b$} \\
\noalign{\smallskip}
 $\geq$Sa \& $Def_{HI}<$0.5	     &  $i$  &  156  &   0.58$\pm$0.04  &  -4.9$\pm$0.2  & 0.64    &  0.17	      \\
				     &  NUV  &  156  &   0.66$\pm$0.04  &  -5.6$\pm$0.1  & 0.72    &  0.17      \\
				     &  FUV  &  147  &   0.67$\pm$0.04  &  -5.7$\pm$0.2  & 0.74    &  0.16     \\

\noalign{\smallskip}							     								 
\noalign{\smallskip}					        
\hline
\hline
\end{tabular}
\end{center}
}
\end{table}

\end{document}